\documentclass[aps,twocolumn,showpacs,preprintnumbers,superscriptaddress,floatfix,nofootinbib]{revtex4-1}

\usepackage[utf8]{inputenc}
\usepackage{slashed}
\usepackage{soul}
\usepackage{empheq}
\usepackage{indentfirst}
\usepackage{overpic}
\usepackage{color}

\usepackage{amsmath,amssymb,amsfonts}
\usepackage{mathrsfs}
\usepackage{stmaryrd,latexsym,bm,bbm}
\usepackage{epsfig,graphicx}
\usepackage{epstopdf}

\usepackage[%dvipdfm,
colorlinks=true,
linkcolor=black,
breaklinks=true,
urlcolor=blue,
citecolor=green]{hyperref}

\usepackage{booktabs}
\usepackage{dcolumn}
\usepackage{makecell}
\usepackage{multirow}

\definecolor{darkblue}{rgb}{0.,0.,0.7}
\definecolor{light-blue}{rgb}{0.8,0.85,1}
\definecolor{green}{rgb}{0,0.6,0}
\definecolor{blueviolet}{rgb}{0.541, 0.169, 0.886}
\definecolor{fuchsia}{rgb}{1.0, 0, 1.0}

\newcommand{\Lag}{\mathcal{L}}

\newcommand{\mM}{\mathcal{M}}

\newcommand{\mev}{\mathrm{MeV}}
\newcommand{\gev}{\mathrm{GeV}}

\begin{document}

\title{Strong decays of the latest LHCb pentaquark candidates in hadronic molecule pictures}

\author{Yong-Hui Lin}\email{linyonghui@mail.itp.ac.cn}
\affiliation{CAS Key Laboratory of Theoretical Physics, Institute
	of Theoretical Physics,  Chinese Academy of Sciences, Beijing 100190, China}
\affiliation{University of Chinese Academy of Sciences (UCAS), Beijing 100049, China}

\author{Bing-Song Zou}\email{zoubs@mail.itp.ac.cn}
\affiliation{CAS Key Laboratory of Theoretical Physics, Institute
	of Theoretical Physics,  Chinese Academy of Sciences, Beijing 100190, China}
\affiliation{University of Chinese Academy of Sciences (UCAS), Beijing 100049, China}
\affiliation{Synergetic Innovation Center for Quantum Effects and Applications (SICQEA),
	 Hunan Normal University, Changsha 410081, China}

\begin{abstract}
  We investigate the observed pentaquark candidates $P_c(4312)$, $P_c(4440)$ and $P_c(4457)$ from the latest LHCb measurement, as well as four possible spin partners in the $\bar{D}^{(*)}\Sigma_c^*$ system predicted from the heavy quark spin symmetry with the hadronic molecule scenarios. Similar to the previous calculation on $P_c(4380)$ and $P_c(4450)$, the partial widths of all the allowed decay channels for these $P_c$ states are estimated with the effective Lagrangian method. The cutoff dependence of our numerical results are also presented. Comparing with the experimental widths, our results show that $P_c(4312)$, $P_c(4440)$ and $P_c(4457)$ can be described well with the spin-parity-$1/2^-$-$\bar{D}\Sigma_c$, $1/2^-$-$\bar{D}^*\Sigma_c$ and $3/2^-$-$\bar{D}^*\Sigma_c$ molecule pictures, respectively.

%  \begin{keywords}
%	~Keywords: unquenched quark model,modified T operator,mass
%  \end{keywords}

\end{abstract}

%\pacs{14.40.Be,12.39.Pn,24.10.Eq}

\maketitle

\section{Introduction}~\label{sec:1}
The latest LHCb measurement observed more precise line shape of the $J/\psi p$ invariant mass distribution from the process $\Lambda_b^0\to J/\psi p K^-$~\cite{Aaij:2019vzc}. The experimental data suggested that the previous observed structure $P_c(4450)$ is resolved into two narrow states, $P_c(4440)$ and $P_c(4457)$ while the broad state $P_c(4380)$ have not been confirmed yet. In addition, a new structure $P_c(4312)$ is discovered with $7.3\sigma$ significance. Their masses and widths are given in the following table.
%---------
\begin{table}[htpb]
\centering
\scalebox{1}{
	\begin{tabular}{*{3}{c}}
			\Xhline{0.8pt}
			States & Mass ($\mev$)& Width ($\mev$)\\
			\Xhline{0.4pt}
			$P_c(4312)^+$ & $4311.9\pm 0.7^{+6.8}_{-0.6}$ & $9.8\pm2.7^{+3.7}_{-4.5}$ 	\\
			\Xhline{0.4pt}
			$P_c(4440)^+$ & $4440.3\pm 1.3^{+4.1}_{-4.7}$ & $20.6\pm4.9^{+8.7}_{-10.1}$ 	\\
			\Xhline{0.4pt}
			$P_c(4457)^+$ & $4457.3\pm 0.6^{+4.1}_{-1.7}$ & $6.4\pm2.0^{+5.7}_{-1.9}$ 	\\
			\Xhline{0.8pt}	
		\end{tabular}
	}
\end{table}
%---------

The reported masses of $P_c(4312)$ and $P_c(4457)$ lie approximately $10\ \mev$ and $5\ \mev$ below the $\bar{D}\Sigma_c$ and $\bar{D}^*\Sigma_c$ thresholds, respectively. This closeness to the thresholds and their narrow widths make the interpretation of hadronic molecule consisting of the corresponding meson-baryon system naturally for these pentaquark-like states. And the experimental properties of previous $P_c(4380)$ and $P_c(4450)$ can be described well in the similar scenarios within some reasonable parameter range~\cite{Lin:2017mtz}. Actually, before the first observation of pentaquark structure in hidden charm sector by LHCb in 2015~\cite{Aaij:2015tga}, the existence of such near threshold bound states has been predicted systematically in some early theoretical works~\cite{Wu:2010jy,Wu:2010vk,Wang:2011rga,Yang:2011wz,Wu:2012md,Yuan:2012wz,Xiao:2013yca}. Especially, the predicted masses for these three observed $P_c$ states in Ref.~\cite{Wu:2012md} are exactly consistent with the reported experimental measurement within the uncertainty. And from the theoretical analysis in that work, we note that the $\bar{D}\Sigma_c$ and $\bar{D}^*\Sigma_c$ account for a large proportion of component in lower $P_c(4312)$ and higher two $P_c$ states, respectively. After that experimental discovery, various other theoretical scenarios have been also proposed to understand the nature of pentaquark-like states, which include compact pentaquarks~\cite{Jaffe:2003sg,Yuan:2012wz,Ali:2016dkf,Maiani:2015vwa,Li:2015gta,Wang:2015epa,Weng:2019ynv,Stancu:2019qga,Giannuzzi:2019esi,Zhu:2019iwm,An:2019idk}, baryocharmonia~\cite{Kubarovsky:2015aaa,Eides:2019tgv} and rescattering-induced kinematical effects~\cite{Guo:2015umn,Liu:2015fea,Guo:2016bkl,Bayar:2016ftu}, as well as other possible bounded mechanism~\cite{Mironov:2015ica,Scoccola:2015nia}. The definite conclusion on the inner structures of $P_c$ states, however, requires further experimental investigation for them, especially the determination of their spin and parity.

Recently, starting off with the near threshold properties of the reported $P_c$ states, some theoretical works suggested the molecular interpretations are favorable to them~\cite{Chen:2019bip,Chen:2019asm,Guo:2019fdo,Liu:2019tjn,He:2019ify,Liu:2019zoy,Huang:2019jlf,Shimizu:2019ptd,Guo:2019kdc,Xiao:2019aya,Xiao:2019mst,Sakai:2019qph}. And additional four similar hadronic molecules are expected with the heavy quark spin symmetry~\cite{Liu:2019tjn,Sakai:2019qph}. The systemic introduction to the hadronic molecules can refer to the reviews~\cite{Chen:2016qju,Guo:2017jvc}. In the present work, we would like to investigate the decay properties of the newly observed $P_c$ states within the $S$-wave hadronic molecular pictures. The strong interactions among the involved hadrons are described with the effective Lagrangian method. As a result, the whole strong decay patterns are presented with the free parameters fixed to reproduce the measured total decay widths. It will help us to verify whether $P_c(4312)$, $P_c(4440)$ and $P_c(4457)$ are $S$-wave hadronic molecule states or not in future. Besides that, another four possible molecules in $\bar{D}^{(*)}\Sigma_c^*$ system predicted in Refs.~\cite{Xiao:2013yca,Liu:2019tjn} are also investigated.

This work is organized as follows: In Sec.~\ref{sec:2}, we introduce formalism and some details about the theoretical tools used to calculate the decay modes of exotic hadronic molecular states. In Sec.~\ref{sec:3}, the numerical results and discussion are presented. The last section is devoted to the summary of the present work.

\section{Formalism}~\label{sec:2}

%\begin{widetext}
%\end{widetext}
%\subsection{Modified $^3P_0$ opertaor}
\subsection{Decay channels}
Since there is no definite experimental evidence to identify the quantum numbers for all of the observed $P_c$ states up to now, we decipher them as the $S$-wave hadronic molecules in the present work. It indicates that $P_c(4312)$ is treated as a $J^P=1/2^-$ $\bar D\Sigma_c$ bound state while $P_c(4440)$ and $P_c(4457)$ are $\bar D^*\Sigma_c$ bound states with two possible quantum numbers $1/2^-$ and $3/2^-$. With the effective Lagrangian approach, the partial decay widths of $P_c$ molecules to all possible channels can be estimated consistently.

\begin{figure}[htbp]
	\begin{center}
		\includegraphics[width=9cm]{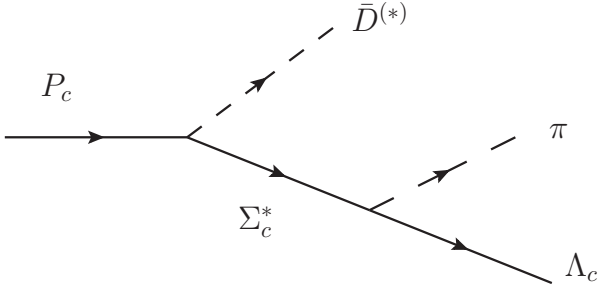}
		\caption{Three-body decays of the $\bar D^{(*)}\Sigma_c^*$ molecules.\label{Fig:three-body}}
	\end{center}
\end{figure}

\begin{figure}[htbp]
	\begin{center}
		\includegraphics{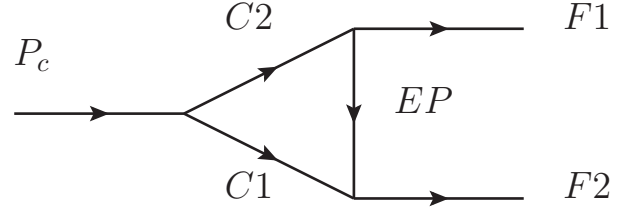}
		\caption{The triangle diagram for the two-body decays of the $P_c$ states in the $\bar D^{(*)}\Sigma_c^{(*)}$ molecule scenarios, where $C1$, $C2$ denote the constituent particles of the $\bar D^{(*)}\Sigma_c^{(*)}$ composite system, $F1$, $F2$ denote the final states, $EP$ denotes the exchanged mesons.\label{Fig:triangle}}
	\end{center}
\end{figure}

Compared with the reported total widths of $P_c$ states, only the effect from the finite width of $\Sigma_c^{*}$~($\sim15~\mev$) needs to be considered and all other constituent hadrons, which include $\bar D$, $\bar D^{*}$ and $\Sigma_c$, can be treated as stable particles. And the natural three-body decays through the bounded $\Sigma_c^*$ decay will contribute to the widths of $\bar D^{(*)}\Sigma_c^*$ molecules, as shown in Fig.~\ref{Fig:three-body}. The two-body decays of hadronic molecules will be described conventionally by the triangle diagram mechanism with the one meson exchanged as in Fig.~\ref{Fig:triangle}.
%---------
%---------
%---------
%---------
All the two-body decay channels considered in our calculation are collected in Table~\ref{Tab:modes}.
%---------
\begin{table}[htpb]
	\centering
	\caption{\label{Tab:modes}All possible decay channels for the $P_c$ states in the $\bar D^{(*)}\Sigma_c^{(*)}$ molecule scenario.}
	 \scalebox{0.8}{
	\begin{tabular}{c|*{2}{c}}
		\Xhline{1.0pt}
		\thead{Initial state} & \thead{Final states} & \thead{Exchanged particles} \\
		\Xhline{0.8pt}
		\multirow{5}*{$P_c(4312)(\bar D \Sigma_c)$} & $J/\psi N$, $\omega p$, $\rho N$ & $D$, $D^*$ \\
		\Xcline{2-3}{0.4pt}
		& $\bar D^*\Lambda_c$ & $\pi$, $\rho$ \\
		\Xcline{2-3}{0.4pt}
		& $\bar D \Lambda_c$ & $\rho$ \\
		\Xcline{2-3}{0.4pt}
		& $\eta_c N$ & $D^*$ \\
		\Xcline{2-3}{0.4pt}
		& $\pi N$ & $D^*$, $\Lambda_c$, $\Sigma_c$ \\
		\Xhline{0.8pt}
		\multirow{4}*{\thead{$P_c(4440)\& P_c(4457)$ \\ $(\bar D^* \Sigma_c)$} } & $\bar D^*\Lambda_c$, $\bar D\Lambda_c$, $\bar D \Sigma_c^*$, $\bar D\Sigma_c$ & $\pi$, $\rho$ \\
		\Xcline{2-3}{0.4pt}
		& $J/\psi N$, $\omega p$, $\rho N$, $\eta N$ & $D^*$, $D$ \\
		\Xcline{2-3}{0.4pt}
		& $\pi N$ & $D^*$, $D$, $\Lambda_c$, $\Sigma_c$ \\
		\Xcline{2-3}{0.4pt}
		& $\chi_{c0} N$ & $D^*$ \\
		\Xhline{0.8pt}
		\multirow{6}*{$P_c(4376)(\bar D \Sigma_c^*)$} & $\bar D^*\Lambda_c$ & $\pi$, $\rho$ \\
		\Xcline{2-3}{0.4pt}
		& $\bar D\Lambda_c$, $\bar D\Sigma_c$ & $\rho$ \\
		\Xcline{2-3}{0.4pt}
		& $J/\psi N$, $\omega p$, $\rho N$ & $D^*$, $D$ \\
		\Xcline{2-3}{0.4pt}
		& $\eta_c N$ & $D^*$ \\
		\Xcline{2-3}{0.4pt}
		& $\pi N$ & $D^*$, $\Lambda_c$, $\Sigma_c$ \\
		\Xcline{2-3}{0.4pt}
		& $\chi_{c0} N$ & $D$ \\
		\Xhline{0.8pt}
		\multirow{4}*{\thead{$P_c(4500)\& P_c(4511)$ \\ \&$P_c(4523)(\bar D^* \Sigma_c^*)$} } & $\bar D^*\Lambda_c$, $\bar D\Lambda_c$, $\bar D \Sigma_c^*$, $\bar D\Sigma_c$, $\bar D\Sigma_c^*$ & $\pi$, $\rho$ \\
		\Xcline{2-3}{0.4pt}
		& $J/\psi N$, $\omega p$, $\rho N$, $\eta N$ & $D^*$, $D$ \\
		\Xcline{2-3}{0.4pt}
		& $\pi N$ & $D^*$, $D$, $\Lambda_c$, $\Sigma_c$ \\
		\Xcline{2-3}{0.4pt}
		& $\chi_{c0} N$ & $D^*$ \\
		\Xhline{1.0pt}
	\end{tabular}
	 }
\end{table}
%---------
\subsection{Effective Lagrangian}
In the present work, we adopt the effective Lagrangian approach to compute the amplitudes of above decay diagrams. For the first vertex that $P_c$ states couple to the hadronic baryon-meson pairs, the Lorentz covariant $L$-$S$ scheme proposed in Ref.~\cite{Zou:2002yy} is used. A remarkable feature of this configuration is that the $L$-$S$ effective Lagrangian contains definite angular momentum contribution of the final two-body system in the decay process. In our $S$-wave molecule scenario, the involved Lagrangian is presented in the following,
\begin{align}
	\Lag_{\bar{D} \Sigma_c P_c(1/2^-)} &= g_{\bar{D} \Sigma_c P_c}^{1/2^-} \bar{\Sigma}_c P_{c} \bar{D}, \\
	\Lag_{\bar{D} \Sigma_c^* P_c(3/2^-)} &= g_{\bar{D} \Sigma_c^* P_c}^{3/2^-} \bar{\Sigma}_c^{* \mu}P_{c \mu} \bar{D}, \\	
	\Lag_{\bar{D}^* \Sigma_c P_c(1/2^-)} &= g_{\bar{D}^* \Sigma_c P_c}^{1/2^-} \bar{\Sigma}_c \gamma^5\tilde{\gamma}^{\mu}P_{c} \bar{D}^{*}_{\mu}, \\
	\Lag_{\bar{D}^* \Sigma_c P_c(3/2^-)} &= g_{\bar{D}^* \Sigma_c P_c}^{3/2^-} \bar{\Sigma}_c P_{c \mu} \bar{D}^{* \mu}, \\
	\Lag_{\bar{D}^* \Sigma_c^* P_c(1/2^-)} &= g_{\bar{D}^* \Sigma_c^* P_c}^{1/2^-} \bar{\Sigma}_c^{* \mu} P_{c} \bar{D}^{*}_{\mu}, \\
	\Lag_{\bar{D}^* \Sigma_c^* P_c(3/2^-)} &= g_{\bar{D}^* \Sigma_c^* P_c}^{3/2^-} \bar{\Sigma}_c^{* \mu}\gamma^5\tilde{\gamma}^{\nu} P_{c \mu} \bar{D}^{*}_{\nu}, \\
	\Lag_{\bar{D}^* \Sigma_c^* P_c(5/2^-)} &= g_{\bar{D}^* \Sigma_c^* P_c}^{5/2^-} \bar{\Sigma}_c^{* \mu} P_{c \mu\nu} \bar{D}^{* \nu},		
	\label{eq:vertex0}
\end{align}
with $\tilde{\gamma}^\mu$ defined as $(g^{\mu\nu}-p^\mu p^\nu/p^2)\gamma_\nu\equiv\tilde{g}^{\mu\nu}\gamma_\nu$, where $p$ denotes the momentum of initial $P_c$ state. The effective couplings $g_{\bar{D}^{(*)}\Sigma_c^{(*)}P_c}$ can be estimated with the compositeness criterion which states the relation between the derivative of self-energy operator of hadron resonance and its compositeness~\cite{Weinberg:1962hj,Weinberg:1965zz}. And the pure $\bar D^{(*)}\Sigma_c^{(*)}$ molecular structures are assumed for $P_c$ states which indicates the compositeness of $P_c$ states equals to one in this work, that is $\chi\equiv1-Z=1$. Working in the non-relativistic limit and expanding on the small account $\sqrt{2\mu E_B}/\Lambda$, the simplest estimation, denoted as $g_0$, for $g_{\bar{D}^{(*)}\Sigma_c^{(*)}P_c}$ can be obtained with only the leading term kept. It is
\begin{align}
g_0&=\sqrt{\frac{8\sqrt{2}\sqrt{E_B}m_1 m_2 \pi}{(m_1 m_2/(m_1+m_2))^{3/2}}}\sqrt{\frac{1}{\mM_{N}F_T}}\label{eq:coupling}\\
F_T&=\begin{cases}
1& \text{for spin-$1/2$ molecule},\\
3/2& \text{for spin-$3/2$ molecule},\\
5/3& \text{for spin-$5/2$ molecule}.
\end{cases},\notag\\
\mM_{N}&=\begin{cases}
2\, m_1& \text{for spin-$1/2$ $\bar{D}\Sigma_c$ molecule},\\
6\,m_1& \text{for spin-$1/2$ $\bar{D}^*\Sigma_c$ molecule},\\
4/3\,m_1& \text{for spin-$3/2$ $\bar{D}\Sigma_c^*$ or $\bar{D}^*\Sigma_c$  molecule},\\
4\,m_1& \text{for spin-$1/2$ $\bar{D}^*\Sigma_c^*$ molecule},\\
20/9\, m_1& \text{for spin-$3/2$ $\bar{D}^*\Sigma_c^*$ molecule},\\
6/5\, m_1& \text{for spin-$5/2$ $\bar{D}^*\Sigma_c$ molecule}.
\end{cases}.\notag
\end{align}
As for the additional Lagrangians required to construct the one meson exchanged potential, we adopt the conventional formula used in a variety of phenomenological approaches. The specific formalism can refer to our previous work~\cite{Lin:2017mtz}. And these effective coupling constants have been organized consistently based on $SU(3)$ flavor symmetry in Refs.~\cite{deSwart:1963pdg,Polinder:2006zh,Ronchen:2012eg,Haidenbauer:2016pva,Haidenbauer:2017sws}. We take the same convention as in Ref.~\cite{Ronchen:2012eg} and extend to get whole coupling relations. In our hidden charm cases, the coupling constants between charmonium and charmed mesons are related to the couplings $ g_1 $, $ g_2 $, respectively, using the heavy quark symmetry~\cite{Colangelo:2003sa,Guo:2010ak}, where $g_1 $ and $ g_2 $, which can be related to the decay constants of $ \chi_{c0} $ and $ J/\psi $ by using the vecrtor-meson-dominance(VMD) arguments\footnote{Note that there is a factor 2 difference for these values in Ref.~\cite{Colangelo:2003sa} and~\cite{Guo:2010ak} due to the difference in conventions.}, are the couplings of the $ P $- and $S$-wave charmonium fields to the charmed and anti-charmed mesons, respectively. In the present calculation, we take the same convention as Ref.~\cite{Guo:2010ak}, that is, $g_1 = -5.4 \ \mathrm{GeV}^{-1/2}$ and $g_2 =  2.1 \ \mathrm{GeV}^{-3/2}$. And the couplings between charmed mesons and light vector mesons can be estimated with the VMD approach~\cite{Lin:1999ad,Oh:2000qr}. Note that the coupling $g_{D^{(*)}D^{(*)}J/\psi}$ is included in both of these two determinations, $g_{DDJ/\psi}=g_{D^*D^*J/\psi}=7.44$, $g_{D^*DJ/\psi}=7.91\ \gev^{-1}$ in VMD, while with heavy quark symmetry, one obtain $g_{DDJ/\psi}=6.95$, $g_{D^*D^*J/\psi}=7.48$, and $g_{D^*DJ/\psi}=7.21\ \gev^{-1}$ (Note that the different values for $g_{DDJ/\psi}$ and $g_{D^*D^*J/\psi}$ is because the experimental masses of $D$ and $D^*$ are used). Since there is no significant difference between these two methods, we take the value of coupling $g_{D^{(*)}D^{(*)}J/\psi}$ in VMD determination. For the effective couplings which have charmed baryon($\Sigma_c^{(*)}, \Lambda_c$) involved, the heavy quark spin symmetry(HQSS) can be applied to reduce the number of undetermined couplings in this part~\cite{Yan:1992gz,Cheng:2004ru}. And the left unknown couplings are estimated by taking the simplest approximation, that is, we assume that the role of charm quark is the same as that of strange quark. In this way, we use the same value from the $SU(3)$ relations, for example, $g_{\rho\Sigma_c\Lambda_c}=g_{\rho\Sigma\Lambda}$. Finally, there is another set of couplings, which includes $g_{D^*D\pi}$, $\pi\Sigma_c\Lambda_c$ and $\pi\Sigma_c^*\Lambda_c$, is inferred from the experimental decay widths. All effective couplings we used are listed in Table~\ref{table:constants}. One should note that most of these values can only be regarded as rough estimations, which should suffice for an order-of-magnitude estimate of the decay rates under consideration.
%---------
\begin{table*}[htpb]
	\centering
	\caption{\label{table:constants}Coupling
		constants used in the present work. The $P$, $V$, $B$ and $D$ denote the pseudoscalar,
		vector mesons, octet and decuplet baryons respectively. Only absolute values of the couplings are listed with
		their signs ignored.}
	\scalebox{1}{
		\begin{tabular}{*{11}{c}}
			\Xhline{1.0pt}
			$\alpha_{BBP}$ & $\alpha_{BBV}$ & $g_{BBP}$ & $g_{BBV}$ & $g_{VPP}$
			& \thead{$g_{VVP}$ \\ ($\mathrm{GeV}^{-1}$)} & \thead{$g_{PBD}$ \\ ($\mathrm{GeV}^{-1}$)} & \thead{$g_{VBD}$ \\
				($\mathrm{GeV}^{-1}$)}  & \thead{$g_{PDD}$ \\ ($\mathrm{GeV}^{-1}$)} & $g_{VDD}$ & $\kappa_{VDD}$ 	\\
			\Xhline{0.4pt}
			0.4 & 1.15 & 13.5 & 3.25 & 3.02 & 12.84 & 15.19 & 20.68 & 12.71 & 7.67 & 6.1 	\\
			\Xhline{0.8pt}
			\thead{$g_{\pi\Sigma_c\Sigma_c}$\\($g_{BBP}$)} & \thead{$g_{DN\Sigma_c}$\\($g_{BBP}$)} & \thead{$g_{DN\Lambda_c}$\\($g_{BBP}$)} & \thead{$g_{\rho\Sigma_c\Sigma_c}$\\($g_{BBV}$)} & \thead{$g_{\rho\Sigma_c\Lambda_c}$\\($g_{BBV}$)} & \thead{$g_{D^*N\Sigma_c}$\\($g_{BBV}$)}& \thead{$g_{D^*N\Lambda_c}$\\($g_{BBV}$)} & \thead{$g_{D^*N\Sigma_c^*}$\\($g_{VBD}$)}& \thead{$g_{DN\Sigma_c^*}$\\($g_{PBD}$)}&\thead{$g_{D^*D^*\eta_c}$ \\ ($\mathrm{GeV}^{-1}$)}&$g_{D^*D\eta_c}$ \\
			\Xhline{0.4pt}
			$2\alpha_{BBP}$ & $1-2\alpha_{BBP}$ & $\frac{1+2\alpha_{BBP}}{\sqrt{3}}$ & $2\alpha_{BBV}$ & $\frac{2(1-\alpha_{BBV})}{\sqrt{3}}$ & $1-2\alpha_{BBV}$ & $\frac{1+2\alpha_{BBV}}{\sqrt{3}}$ & $\frac{1}{\sqrt{6}}$ & $\frac{1}{\sqrt{6}}$ & 3.52& 6.82	\\
			\Xhline{0.8pt}
			$g_{\pi\Lambda_c\Sigma_c}$ & \thead{$g_{\pi\Lambda_c\Sigma_c^*}$ \\ ($\mathrm{GeV}^{-1}$)} & $g_{D^*D\pi}$
			& \thead{$g_{D^*D^*\pi}$\footnote{$g_{D^*D^*\pi}$ is related to $g_{D^*D\pi}$ with HQSS, that is, $g_{D^*D^*\pi}=2g_{D^*D\pi}/\sqrt{m_{D^*}m_{D}}$. Note that compared with that in Ref.~\cite{Cheng:2004ru}, an additional factor 2 is included duo to the different Lagrangian for the $D^*D\pi$ interaction we used here. And the value of $g_{D^*D\pi}$ is a factor of $\sqrt{2}$ smaller than that in Ref.~\cite{Lin:2017mtz} due to the difference in conventions.} \\ ($\mathrm{GeV}^{-1}$)}& \thead{$g_{D^*D\rho}$ \\ ($\mathrm{GeV}^{-1}$)}& $g_{D^*D^*\rho}$& $g_{DD\rho}$& \thead{$g_{D^*D\omega}$ \\ ($\mathrm{GeV}^{-1}$)}& $g_{D^*D^*\omega}$& $g_{DD\omega}$&\thead{$g_{D^*DJ/\psi}$ \\ ($\mathrm{GeV}^{-1}$)} \\
			\Xhline{0.4pt}
			19.31 & 7.46 & 6.0 & 6.2 & 2.51 & 2.52 & 2.52 & 2.83 & 2.84 & 2.84 &7.94	\\
			\Xhline{0.8pt}
			$g_{D^*D^*J/\psi}$& $g_{DDJ/\psi}$& $g_{DD\chi_{c0}}$&\thead{$g_{D^*D^*\chi_{c0}}$ \\ ($\mathrm{GeV}^{-1}$)} \\
			\Xhline{0.4pt}
			7.44 & 7.44 & 32.24 & 11.57 	\\
			\Xhline{1.0pt}	
		\end{tabular}
	}
\end{table*}
%---------
\subsection{Form factors}
As discussed in our previous work, some of the triangle diagrams, corresponding to the exchange of a pseudoscalar meson for the $D$-wave decay modes~\cite{Albaladejo:2015dsa,Shen:2016tzq}, are ultraviolet(UV) finite while the others diverge when the UV finite loops receive short-distance contributions if we integrate over the whole momentum space. We will employ the following UV regulator which suppress short-distance contributions and thus can render all the amplitudes UV finite~\cite{Faessler:2007gv,Dong:2009yp,Dong:2009tg,Lu:2016nnt,Xiao:2019mst}
\begin{equation}
f_1(p^2_E /\Lambda_0^2) = {\rm{exp}}(-p^2_E /\Lambda_0^2),
\label{eq:regulator4}
\end{equation}
where $p_E$, defined as
${m_{\bar{D}^{(*)}}}p_{\Sigma_c^{(*)}}/({m_{\bar{D}^{(*)}}+m_{\Sigma_c^{(*)}}})-
{m_{\Sigma_c^{(*)}}}p_{\bar{D}^{(*)}}/({m_{\bar{D}^{(*)}}+m_{\Sigma_c^{(*)}}}) $ for the $
\bar{D}^{(*)}\Sigma_c^{(*)}$ molecules,  is the Euclidean Jacobi momentum. The cutoff $\Lambda_0$ denotes a hard momentum scale which suppresses the contribution of the two constituents at short distances $\sim 1/\Lambda_0$. There is no universal criterion for the determination of these cut-offs and even for the choice of the regulator functions, but as a general rule the value of $\Lambda_0$ should be much larger than the typical momentum in the bound state, given by $\sqrt{2\mu\epsilon}$ ($\sim 0.1\ \gev$ for the $P_c$ molecules). And it should also not be too large since we have neglected all other degrees of freedom, except for the two constituents, which would play a role at short distances. In the present work, we vary the value of $\Lambda_0$ from $0.6\ \mathrm{GeV}$ to $1.4\ \mathrm{GeV}$ for a rough estimate of the two-body partial widths. Note that there is another three-momentum Gaussian form factor is routinely used in a variety of non-relativistic phenomenological approaches~\cite{Nieves:2012tt,HidalgoDuque:2012pq,Guo:2017jvc},
\begin{equation}
f_2(\bm{p}^2 /\Lambda_0^2) = {\rm{exp}}(-\bm{p}^2 /\Lambda_0^2),
\label{eq:regulator3}
\end{equation}
where $\bm p$ is the spatial part of the momentums of $\bar{D}^{(*)}$ and $\Sigma_c^{(*)}$ in the rest frame of $P_c$ states. The significant difference between these two Gaussian form regulators is that $f_1$ includes an additional constraint on the energy of molecular components, which demands that the center of mass energy is divided as the mass distribution of compounding particles inside the molecular states. It occurs usually for the bound states in quantum mechanics. We will discuss the effect of this energy constraint when we present our numerical results.

In addition, a multipolar form factor is introduced to suppress the off-shell contributions of the exchanged mesons in our triangle diagrams. It is chosen as
\begin{equation}
f_3(q^2) = \frac{\Lambda_1^4}{(m^2 - q^2)^2 + \Lambda_1^4},
\label{eq:multipolar}
\end{equation}
where $m$ and $q$ is the mass and momentum of the exchanged particle. The parameter $\Lambda_1$ is also varied in the range of $0.6$-$1.4 \ \mathrm{GeV}$.

With the effective Lagrangian method, the partial decay widths of $P_c$ states are computed in the perturbative language,
\begin{equation}
{\rm d}\Gamma = \frac{F_I}{32 \pi^2} \overline{|{\cal M}|^2}
\frac{|\mathbf{p_1}|}{M^2} {\rm d}\Omega,
\label{eq:widths}
\end{equation}
where ${\rm d}\Omega = {\rm d}\phi_1 {\rm d}(\cos{\theta_1})$ is the solid angle of the final state in the rest frame of $P_c$, $M$ is the mass of decaying $P_c$ states, the factor $F_I$ is from the isospin symmetry, and the
polarization-averaged squared amplitude $\overline{|{\cal M}|^2}$ means $\frac1{2J+1} \sum_\text{spin} |{\cal M}|^2$\footnote{Since the relative phase between the amplitudes contributed from the different exchanged particles in a specific decay channel can not be determined definitely, we compute the incoherence summation for various decay processes, e.g., $|{\cal M}|^2=|{\cal M_{\pi}}|^2+|{\cal M_{\rho}}|^2$ for $\bar D^*\Lambda_c$ final state.} with $J$ the spin of $P_c$.

\section{Numerical Results and Discussions}~\label{sec:3}
With the effective coupling constants collected, the partial decay widths of observed $P_c$ states can be computed numerically by using the effective Lagrangian approach in the hadronic molecule scenarios. Note that there are still two undetermined parameters in our calculation, $\Lambda_0$ and $\Lambda_1$. The existence of such energy scale parameters is inevitable in the phenomenological paradigms of strong interaction, either introduced to eliminate the loop divergence or to indicate the energy range where the effective approaches do work. As discussed above, we vary these two cut-offs in the range of $0.6$-$1.4\ \gev$ to scrutinize how the decay behaviors undergo changes as the cut-off is varied. And a specific set of values for $\Lambda_0$ and $\Lambda_1$ is chosen to give the decay patterns of $P_c$ molecules by fitting to the measured total widths.

\begin{figure}[htbp]
	\begin{center}
		\includegraphics[width=9cm]{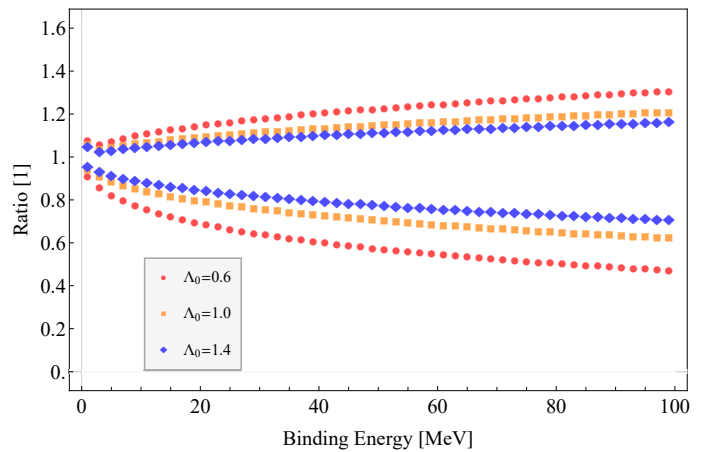}
		\caption{The dependence of relative ratios $g_{RT}/g_{NR}$, $g_0/g_{NR}$ on binding energy, where $g_{NR}$ and $g_{RT}$ denote the non-relativistic and relativistic estimation for the effective couplings between $P_c(4312)$ and $\bar D\Sigma_c$ composite system. And $g_0$ is the approximate $g_{NR}$ as shown in Eq.~\eqref{eq:coupling}. The red-circle, orange-square and blue-diamond dots denote $\Lambda_0=0.6\ \gev$, $1.0\ \gev$ and $1.4\ \gev$, respectively. The upper dots are the values of $g_{RT}/g_{NR}$ while the lower dots are $g_0/g_{NR}$.
		\label{Fig:coupling}}
	\end{center}
\end{figure}

Before going to the discussion on partial decay widths, let us take a moment to figure out the determination of the effective couplings between $P_c$ states and the compounding $\bar D^{(*)}\Sigma_c^{(*)}$ system, $g_{P_c\bar D^{(*)}\Sigma_c^{(*)}}$. As mentioned before, this coupling is estimated with the compositeness condition. It suggests that such a coupling can be expressed as the square root of the inverse of the derivative of its self-energy operator, that has the constituents as the intermediate loop, for a pure molecule state. Since the mass of the hadronic molecule are usually close to the threshold of its constituents, the non-relativistic treatment can be adopted for the estimation of the couplings $g_{P_c\bar D^{(*)}\Sigma_c^{(*)}}$. In Fig.~\ref{Fig:coupling}, we show the differences among three different strategies for the $g_{P_c\bar D\Sigma_c}$ determination, that is, the relativistic calculation denoted as $g_{RT}$, non-relativistic calculation $g_{NR}$ and the $g_0$ which is the approximate estimate of $g_{NR}$ as discussed above. Here, the cutoff $\Lambda_0$ is appeared in the form factor $f_1$ and $f_2$ for removing the UV divergence in the self-energy operator. $f_1$ is related to the relativistic calculation while $f_2$ is used in the non-relativistic case. And with only the leading order left in Eq.~\eqref{eq:coupling}, $g_0$ is cut-off independent. The results show that $g_{RT}$ is always larger than the $g_{NR}$ while $g_0$ is smaller than $g_{NR}$. And as expected, the difference between them increases with the increasing of the binding energy. At the zero-binding-energy limit, the same coupling constant will be obtain from these three various determinations. Since $g_0$ is $\Lambda$-independent, the dependence of $g_{RT}$ and $g_{NR}$ on the cut-off can be translated into the behaviors of these relative ratios change with $\Lambda_0$. As shown in the Fig.~\ref{Fig:coupling}, the lower blue-diamond dot is larger than the lower red-circle dot at the same binding energy which reflects that $g_{NR}$ decreases with the increasing of $\Lambda_0$. And the relative ratio between the upper and lower dot with the same $\Lambda_0$ and binding energy is smaller for the larger cut-off. It means that $g_{RT}$ decreases also as $\Lambda_0$ increases. The cases are similar for the $\bar D^*\Sigma_c$ and $\bar D^{(*)}\Sigma_c^*$ molecule states. Notice that in our molecular scenarios, the binding energy is around $10$, $20$ and $5\ \mev$ for the observed $P_c(4312)$, $P_c(4440)$ and $P_c(4457)$ states respectively. Then there is no significant difference which strategy one adopt for the $g_{P_c\bar D^{(*)}\Sigma_c^{(*)}}$ determination. In the present work, $g_{RT}$ is adopted for these $P_c$ molecules with the binding energy larger than $10\ \mev$. And for $P_c(4312)$, $P_c(4457)$, $P_c(4376)$ and $P_c(4523)$ that have small binding energy, $g_{0}$ is used for simplicity.
%---------
%-----------

\begin{figure*}[htbp]
	\begin{center}
		\includegraphics[width=18cm]{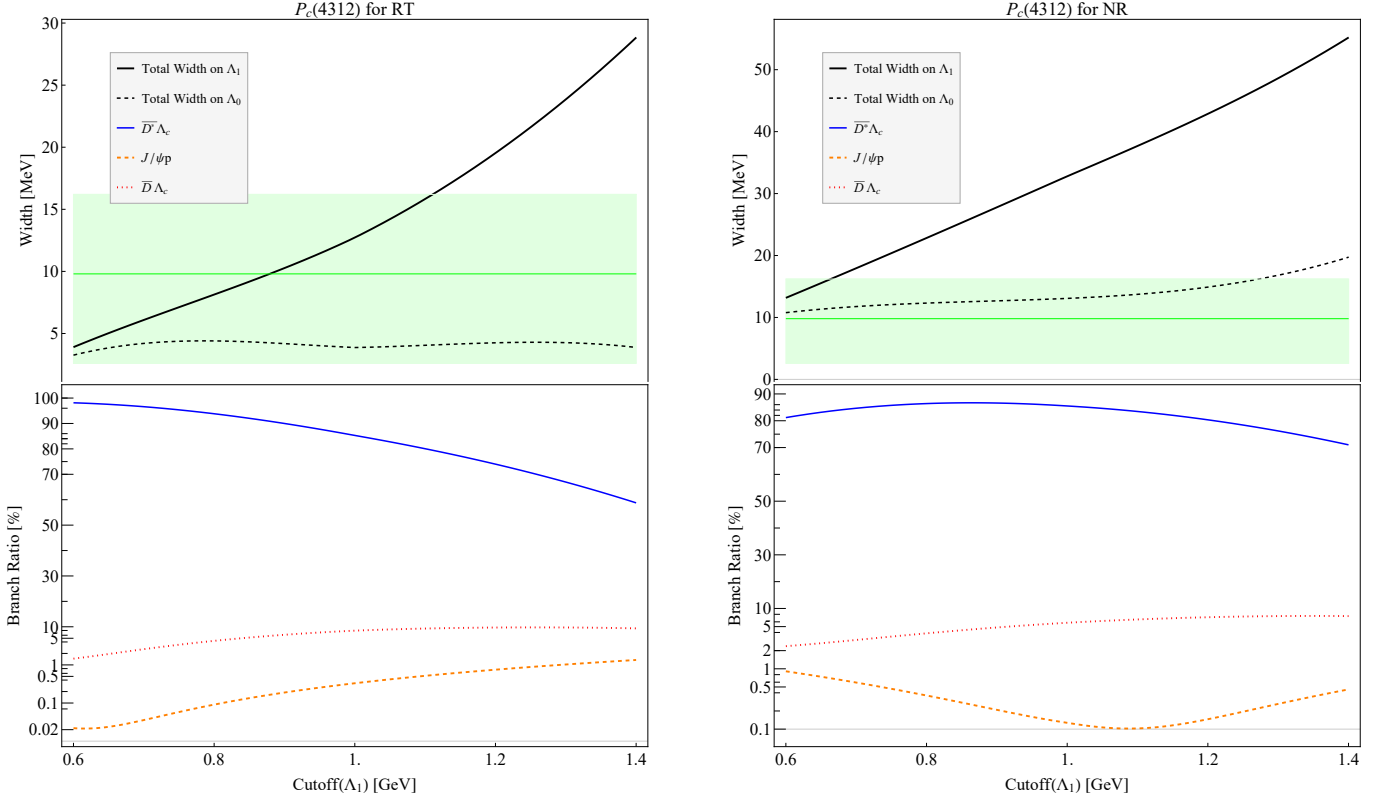}
		\caption{$\Lambda_1$-dependence of the total decay width and the branching fractions of the $\bar D^*\Lambda_c$, $J/\psi p$ and $\bar D\Lambda_c$ channels for the $P_c(4312)$ in the $J^P=1/2^-$ $\bar D \Sigma_c$ molecule scenario. The form factor set is chosen as $(f_1, f_3)$(denoted as RT) for the left panel, and it is $(f_2, f_3)$(denoted as NR) for the right panel. The black solid line denotes the $\Lambda_1$-dependence of total widths while dashed line is the $\Lambda_0$-dependence. And the blue-solid, origin-dashed and red-dotted lines represent the $\Lambda_1$-dependence of partial widths for the $\bar D^*\Lambda_c$, $J/\psi p$ and $\bar D\Lambda_c$ channels, respectively. The green bands in the upper half panels represent the measured widths with uncertainties and the green-solid line denotes the central value.\label{Fig:width-4312}}
	\end{center}
\end{figure*}

\begin{figure*}[htbp]
	\begin{center}
		\includegraphics[width=18cm]{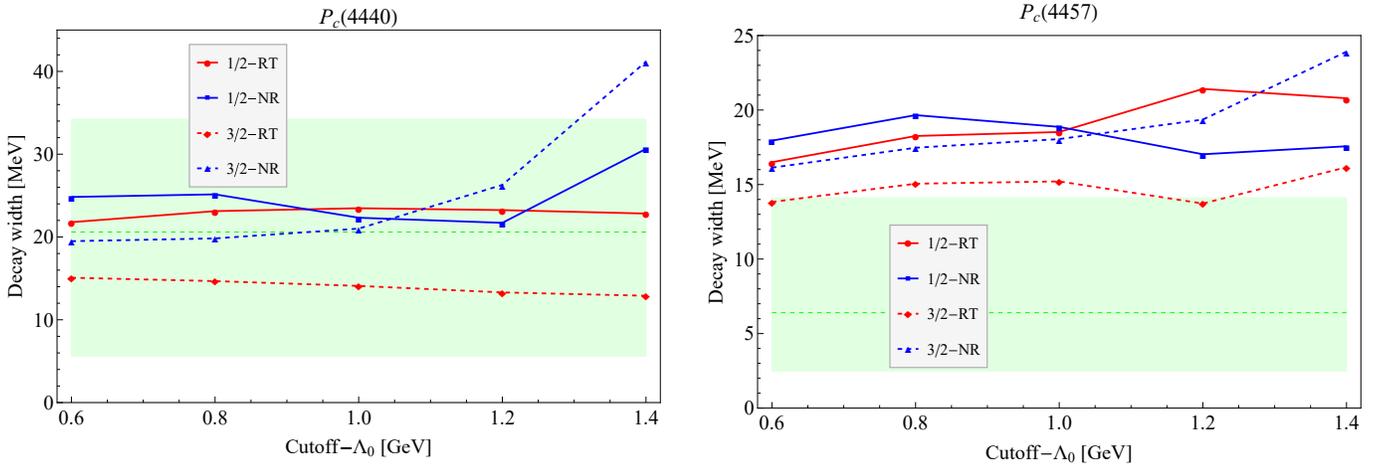}
		\caption{$\Lambda_0$-dependence of the total decay widths for the $P_c(4440)$ in the left panel and $P_c(4457)$ in the right panel, where the blue-solid, blue-dashed, red-solid and red-dashed lines denote that $1/2^-$-$\bar D^*\Sigma_c$ molecule with the form factor set $(f_2, f_3)$, $3/2^-$-$\bar D^*\Sigma_c$ with $(f_2, f_3)$, $1/2^-$-$\bar D^*\Sigma_c$ with $(f_1, f_3)$ and $3/2^-$-$\bar D^*\Sigma_c$ with $(f_1, f_3)$, respectively. The green bands in the upper half panels represent the measured widths with uncertainties and the green-solid line denotes the central value. \label{Fig:width-VB-lambda0}}
	\end{center}
\end{figure*}

\begin{figure*}[htbp]
	\begin{center}
		\includegraphics[width=18cm]{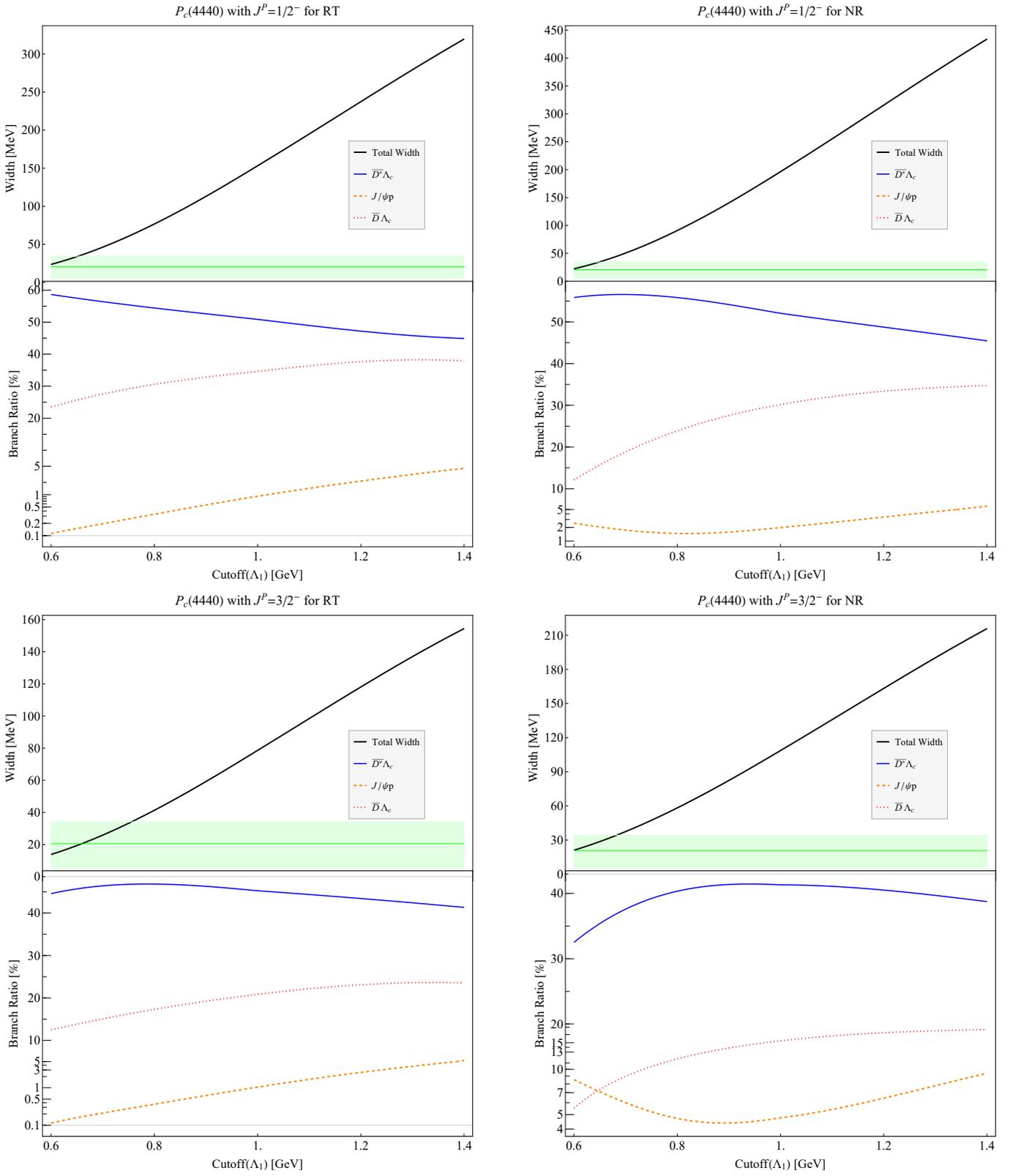}
		\caption{$\Lambda_1$-dependence of the total decay width and the branching fractions of the $\bar D^*\Lambda_c$, $J/\psi p$ and $\bar D\Lambda_c$ channels for the $P_c(4440)$ in the $\bar D^* \Sigma_c$ molecule scenario with $J^P=1/2^-$ or $3/2^-$. The form factor set is chosen as $(f_1, f_3)$(denoted as RT) for the left panel, and it is $(f_2, f_3)$(denoted as NR) for the right panel. The black solid line denotes the $\Lambda_1$-dependence of total widths. And the blue-solid, origin-dashed and red-dotted lines represent the $\Lambda_1$-dependence of partial widths for the $\bar D^*\Lambda_c$, $J/\psi p$ and $\bar D\Lambda_c$ channels, respectively. The green bands in the upper half panels represent the measured widths with uncertainties and the green-solid line denotes the central value.\label{Fig:width-4440}}
	\end{center}
\end{figure*}

\begin{figure*}[htbp]
	\begin{center}
		\includegraphics[width=18cm]{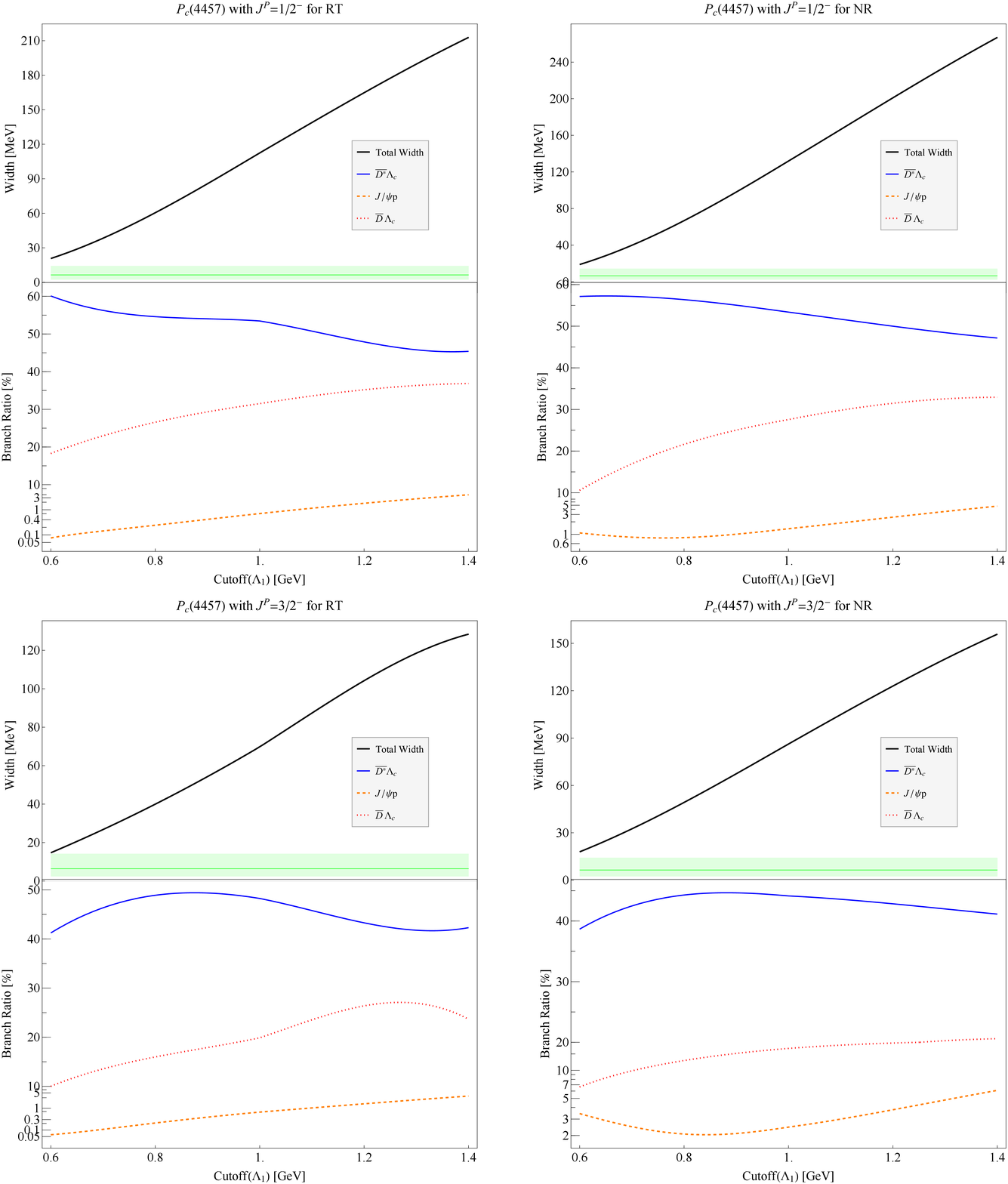}
		\caption{$\Lambda_1$-dependence of the total decay width and the branching fractions of the $\bar D^*\Lambda_c$, $J/\psi p$ and $\bar D\Lambda_c$ channels for the $P_c(4457)$ in the $\bar D^* \Sigma_c$ molecule scenario with $J^P=1/2^-$ or $3/2^-$. The notations are same with Fig.~\ref{Fig:width-4440}.\label{Fig:width-4457}}
	\end{center}
\end{figure*}

The partial decay widths of $P_c(4312)$, $P_c(4440)$ and $P_c(4457)$ in the $S$-wave hadronic molecule pictures with $\Lambda_0=1.0\ \gev$ and $\Lambda_1=0.6\ \gev$ are displayed in Table~\ref{table:widths1} for form factor set ($f_1$, $f_3$) and Table~\ref{table:widths2} for form factor set ($f_2$, $f_3$). And the cutoff-dependence of total widths and the branch fractions of $\bar D^*\Lambda_c$, $J/\psi p$ and $\bar D\Lambda_c$ channels are presented in Fig.~\ref{Fig:width-4312} for $P_c(4312)$ and Fig.~\ref{Fig:width-VB-lambda0}, Fig.~\ref{Fig:width-4440}, as well as Fig.~\ref{Fig:width-4457} for $P_c(4440)$ and $P_c(4457)$ states.
%-----------
\begin{table}[htpb]
	\centering
	\caption{\label{table:widths1}Partial widths of $P_c(4312)$ as $S$-wave $\bar D \Sigma_c$ molecule, $P_c(4440)$ and $P_c(4457)$ as $S$-wave $\bar D^*\Sigma_c$ molecules with two possible quantum numbers, to various possible final states with $\Lambda_0=1.0\ \mathrm{GeV}$, $\Lambda_1=0.6\ \mathrm{GeV}$. The form factor set ($f_1$, $f_3$) is chosen. All of the decay widths are in the unit of $\mathrm{MeV}$, and the short bars denote that this decay channel is closed or the corresponding contribution is negligible. $2\time10^{-4}$ denotes $2\times10^{-4}$.}
	\begin{tabular}{l|*{5}{c}}
		\Xhline{1pt}
		\multirow{3}*{Mode} & \multicolumn{5}{c}{Widths ($\mathrm{MeV}$) with ($f_1$, $f_3$)} \\
		\Xcline{2-6}{0.4pt}
		& \multicolumn{1}{c}{$\bar D \Sigma_c$} & \multicolumn{4}{c}{$\bar D^*\Sigma_c$} \\
		\Xcline{2-2}{0.4pt}\Xcline{3-4}{0.4pt}\Xcline{5-6}{0.4pt}
		& \multicolumn{1}{c}{$P_c(4312)$}& \multicolumn{2}{c}{$P_c(4440)$} & \multicolumn{2}{c}{$P_c(4457)$} \\
		\Xcline{2-2}{0.4pt}\Xcline{3-4}{0.4pt}\Xcline{5-6}{0.4pt}
		& \multicolumn{1}{c}{${\frac12}^-$}& \multicolumn{1}{c}{${\frac12}^-$} & \multicolumn{1}{c}{${\frac32}^-$} & \multicolumn{1}{c}{${\frac12}^-$} & \multicolumn{1}{c}{${\frac32}^-$} \\
		\Xhline{0.8pt}
		$\bar D^*\Lambda_c$ 	 &3.8 &13.9 &6.2 &12.5 &6.1 \\
		$J/\psi p$ 		 		 &0.001 &0.03 &0.02 &0.02 &0.01 \\
		$\bar D\Lambda_c$  	     &0.06 &5.6 &1.7 &3.8 &1.5 \\
		$\pi N$ 			 	 &0.004 &0.002 &$2\time10^{-4}$ &0.001 &$1\time10^{-4}$ \\
		$\chi_{c0}p$ 		 	 &- &$8\time10^{-4}$ &$4\time10^{-5}$ &$9\time10^{-4}$ &$3\time10^{-5}$ \\
		$\eta_c p$ 		 	     &0.01 &$3\time10^{-4}$ &$8\time10^{-5}$ &$2\time10^{-4}$ &$6\time10^{-5}$ \\
		$\rho N$ 			  	 &$3\time10^{-5}$ &$3\time10^{-4}$ &$4\time10^{-5}$ &$2\time10^{-4}$ &$2\time10^{-5}$ \\
		$\omega p$ 		 	     &$1\time10^{-4}$ &0.001 &$2\time10^{-4}$ &$6\time10^{-4}$ &$9\time10^{-5}$ \\
		$\bar D\Sigma_c$ 	  	 &- &3.4 &0.5 &2.6 &1.0 \\
		$\bar D\Sigma^*_c$ 	  	 &- &0.8 &5.4 &1.9 &6.2 \\
		\Xhline{0.8pt}
		Total 				 	 &3.9 &23.7 &13.9 &20.7 &14.7 \\
		\Xhline{1pt}
	\end{tabular}
\end{table}
%-----------
%-----------
\begin{table}[htpb]
	\centering
	\caption{\label{table:widths2}The numerical results for the form factor set ($f_2$, $f_3$). The notation is same with Table~\ref{table:widths1}.}
	\begin{tabular}{l|*{5}{c}}
		\Xhline{1pt}
		\multirow{3}*{Mode} & \multicolumn{5}{c}{Widths ($\mathrm{MeV}$) with ($f_2$, $f_3$)} \\
		\Xcline{2-6}{0.4pt}
		& \multicolumn{1}{c}{$\bar D \Sigma_c$} & \multicolumn{4}{c}{$\bar D^*\Sigma_c$} \\
		\Xcline{2-2}{0.4pt}\Xcline{3-4}{0.4pt}\Xcline{5-6}{0.4pt}
		& \multicolumn{1}{c}{$P_c(4312)$}& \multicolumn{2}{c}{$P_c(4440)$} & \multicolumn{2}{c}{$P_c(4457)$} \\
		\Xcline{2-2}{0.4pt}\Xcline{3-4}{0.4pt}\Xcline{5-6}{0.4pt}
		& \multicolumn{1}{c}{${\frac12}^-$}& \multicolumn{1}{c}{${\frac12}^-$} & \multicolumn{1}{c}{${\frac32}^-$} & \multicolumn{1}{c}{${\frac12}^-$} & \multicolumn{1}{c}{${\frac32}^-$} \\
		\Xhline{0.8pt}
		$\bar D^*\Lambda_c$ 	 &10.7 &12.5 &6.8 &10.8 &6.9 \\
		$J/\psi p$ 		 	     &0.1 &0.6 &1.8 &0.2 &0.6 \\
		$\bar D\Lambda_c$  	     &0.3 &2.7 &1.2 &2.0 &1.2 \\
		$\pi N$ 			 	 &1.7 &0.2 &1.9 &0.07 &0.6 \\
		$\chi_{c0}p$ 		 	 &- &0.1 &0.009 &0.05 &0.003 \\
		$\eta_c p$ 		 	     &0.4 &0.07 &0.008 &0.02 &0.003 \\
		$\rho N$ 			  	 &0.0008 &0.4 &0.3 &0.1 &0.1 \\
		$\omega p$ 		 	     &0.003 &1.5 &1.2 &0.5 &0.4 \\
		$\bar D\Sigma_c$ 	  	 &- &3.4 &0.6 &2.8 &0.9 \\
		$\bar D\Sigma^*_c$ 	  	 &- &0.9 &7.3 &2.3 &7.2 \\
		\Xhline{0.8pt}
		Total 				 	 &13.2 &22.4 &21.0 &18.8 &17.9 \\
		\Xhline{1pt}
	\end{tabular}
\end{table}
%-----------

At first glance, one thing can be concluded that $\bar D^*\Lambda_c$ is the dominant decay channel for both $\bar D\Sigma_c$ and $\bar D^*\Sigma_c$ molecules which is similar with the results on $\bar D\Sigma_c^*$ and $\bar D^*\Sigma_c$ molecules in our previous work~\cite{Lin:2017mtz}. And one can also notice that $\bar D\Lambda_c$ and $\bar D\Sigma_c^{(*)}$ channels also account for a large portion of the widths for the $\bar D^*\Sigma_c$ molecules. In fact, the large partial widths of these channels come from the $\pi$ exchanged contribution. It is because that the exchanged $\pi$ can go nearly on the mass shell in these decay processes. The strong coupling to $\bar D^*\Lambda_c$ channel of $P_c(4312)$ is also claimed in Ref.~\cite{Weng:2019ynv} with the extended chromomagnetic model. And the small $J/\psi p$ decays for all of $S$-wave molecules in our calculation are consistent with the latest LHCb observation which shows that the upper limits of the branching fractions $\mathcal{B}(P_c^+\to J/\psi p)$ are $4.6\%$, $2.3\%$ and $3.8\%$ for $P_c(4312)$, $P_c(4312)$ and $P_c(4312)$ respectively at $90\%$ confidence level by assuming $J^P=3/2^-$ for all of $P_c$ states~\cite{Ali:2019lzf}. And as shown in Refs.~\cite{Weng:2019ynv,Voloshin:2019aut,Sakai:2019qph}, the partial width of $\eta_c p$ channel is almost three times larger than that of $J/\psi p$ for the lowest $P_c(4312)$ state. And the decay width of $P_c(4312)$ to $\bar D\Lambda_c$ is a factor of 0.02 smaller than $\bar D^*\Lambda_c$ channel~\cite{Weng:2019ynv}. These relative ratios are consistent with our calculation as we can see from Table~\ref{table:widths2}. Besides that, our results show that the partial width of $\eta_c p$ channel is around one order of magnitude smaller than that of $J/\psi p$ for the $P_c(4440)$ state. It agrees with the argument of the heavy quark symmetry in Ref.~\cite{Voloshin:2019aut}. Compared with Table~\ref{table:widths1} and Table~\ref{table:widths2}, it does not escape attention that a remarkable difference between the form factor $f_1$ and $f_2$ is that the much larger $D^{(*)}$ meson-exchanged contribution is obtained with $f_2$ when we take the same value of cutoff. According to the definitions of $f_1$ and $f_2$, we know that $f_1$ provides an additional constraint on the energy of compounding particles inside the $P_c$ molecules. Then in that case, the exchanged $D$ or $D^*$ mesons must be highly off the mass shell and this off-shell contribution will be suppressed by our second form factor $f_3$.
%And for the decay of $P_c(4312)$ to $\bar D^*\Lambda$, the partial width obtained with $f_2$ is nearly two times lager than that with $f_1$ while the width of $\bar D^*\Lambda$ channel does not change for the $\bar D^*\Sigma_c$ molecules. As constrained by form factor $f_1$, the dominant integral region of $\bar D\Sigma_c$ to $\bar D^*\Lambda$ is where $\bar D$ and $\Sigma_c$ are nearly close to their mass shell with energy $1.86\ \gev$ and $2.45\ \gev$, respectively. Then the energy carried by the exchanged $\pi$ is around $0.155\ \gev$ which indicates that $\pi$ can also go nearly on the mass shell. While the cases is different for the $\bar D^*\Sigma_c$ molecule where the energy of the off-shell exchanged $\pi$ is around $0.076\ \gev$ in the dominant integral region. In the non-relativistic limit, for the widths obtained with the form factor $f_2$, there is an additional contribution come from the residue which is corresponding to the situation that the exchanged $\pi$ is on its mass shell. Then this extra term is compatible with the width obtained with $f_1$ for the $\bar D\Sigma_c$ molecule since both of them are mainly contributed from the kinematic mechanism where all the intermediate $\bar D$, $\Sigma_c$ and $\pi$ are nearly on their mass shell. For the $\bar D^*\Sigma_c$ molecule, the additional residue, however, indicates that at least one of $\bar D^*$ and $\Sigma_c$ is off mass shell and should be much smaller than the width obtained with $f_1$.
Since the majority of the total widths of $P_c$ molecules is contributed by the $\pi$ exchanged processes which are similar for these two different form factors, the total decay widths of $\bar D\Sigma_c$ and $\bar D^*\Sigma_c$ molecules obtained with $f_1$ and $f_2$ are compatible with each other. And as shown in Fig.~\ref{Fig:width-4312}, Fig.~\ref{Fig:width-4440} and Fig.~\ref{Fig:width-4457}, the cut-off dependence of total widths and branch fractions of $\bar D^*\Lambda_c$, $J/\psi p$ and $\bar D\Lambda_c$ channels is almost same for these two form factors. The total widths increase as $\Lambda_0$ or $\Lambda_1$ increases while the branch fractions are almost stable over the whole range of $\Lambda_1$. It should be noted that $\Lambda_0=1.0\ \gev$ and $\Lambda_1=0.6\ \gev$ are fixed to give a compatible descriptions with measured widths for all of three observed $P_c$ states. The numerical decay patterns with these cutoffs in Table~\ref{table:widths1} suggest that the spin parties of $P_c(4440)$ and $P_c(4457)$ are more likely to be $1/2^-$ and $3/2^-$, respectively. Looking further ahead, the relative ratios between the $\bar D\Sigma_c$ and $\bar D\Sigma_c^*$ and between the $\eta_c p$ and $J/\psi p$ are quite different for the different quantum numbers of $\bar D^*\Sigma_c$ molecules. $\Gamma_{\bar D\Sigma_c}/\Gamma_{\bar D\Sigma_c^*}$ is around 4 for the $1/2^-$-$P_c(4440)$ while it is 0.1 for the $3/2^-$-$P_c(4440)$. And $\Gamma_{J/\psi p}/\Gamma_{\eta p}$ is around 10 for the $1/2^-$-$P_c(4440)$ while it is around 200 for the $3/2^-$-$P_c(4440)$. These novel properties on the branch fractions also exist for the $P_c(4457)$. It will help us to determine the quantum numbers for $P_c(4440)$ and $P_c(4457)$ states experimentally in future.
%-----------
%---------
%---------
%---------
%---------
%---------
%---------
%---------

\begin{figure*}[htbp]
	\begin{center}
		\includegraphics[width=18cm]{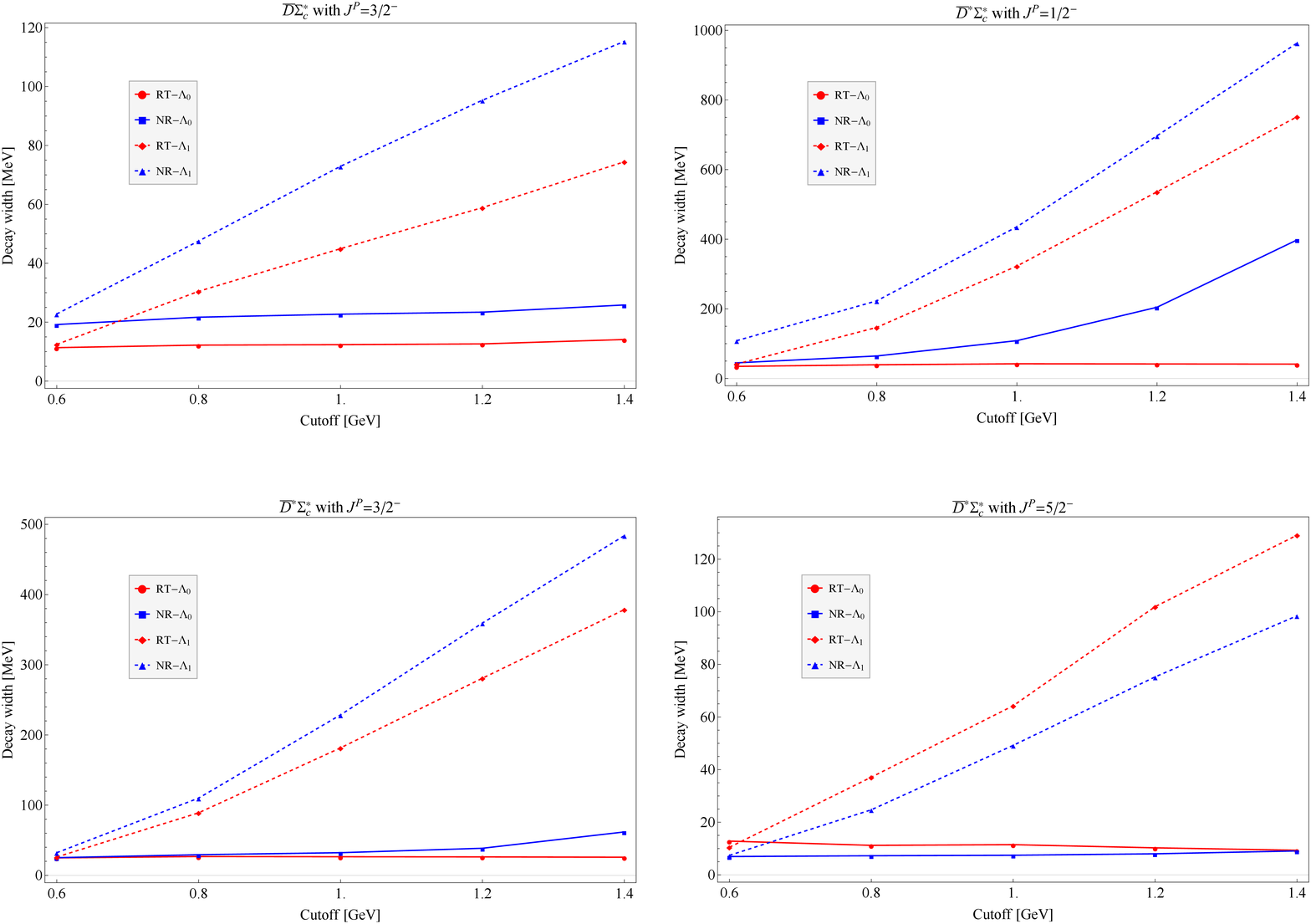}
		\caption{$\Lambda$-dependence of the total decay withs for the four spin partners predicted by Ref.~\cite{Liu:2019tjn} in the $\bar D^{(*)}\Sigma_c^*$ molecule pictures, where the blue-solid, blue-dashed, red-solid and red-dashed lines denote that the dependence on $\Lambda_0$ with form factor set $(f_2, f_3)$, dependence on $\Lambda_1$ with $(f_2, f_3)$, dependence on $\Lambda_0$ with $(f_1, f_3)$ and dependence on $\Lambda_1$ with $(f_1, f_3)$, respectively.\label{Fig:width-VD}}
	\end{center}
\end{figure*}

The quantum numbers of these two $P_c$ states are also discussed with the molecular scenarios in Refs.~\cite{Yamaguchi:2019seo,Valderrama:2019chc,Liu:2019zvb,Pan:2019skd}. And following the heavy quark spin symmetry, Ref.~\cite{Liu:2019tjn} studied all possible heavy quark multiplets in $\bar D\Sigma_c$, $\bar D\Sigma_c^*$, $\bar D^*\Sigma_c$, $\bar D^*\Sigma_c^*$ systems with two sets of quantum numbers for $P_c(4440)$ and $P_c(4457)$ as inputs, ($1/2^-$, $3/2^-$) which they call set A and the opposite identification set B. Since the mass for $1/2^-$-$\bar D\Sigma_c$ molecule produced with set A is more compatible with the LHCb observation, four predicted heavy quark multiplets from the set A are considered in our work, that is, $3/2^-$-$P_c(4376)$, $1/2^-$-$P_c(4500)$, $3/2^-$-$P_c(4511)$ and $5/2^-$-$P_c(4523)$. With the same cutoffs, the partial decay widths of the $S$-wave $\bar D^{(*)}\Sigma_c^*$ molecules are presented in Table~\ref{table:widths3} for form factor set ($f_1$, $f_3$) and Table~\ref{table:widths4} for set ($f_2$, $f_3$). And also the cut-off dependence of total widths are presented in Fig.~\ref{Fig:width-VD}. The decay pattern of $3/2^-$-$\bar D\Sigma_c^*$ molecule is quite the same with the $1/2^-$-$\bar D\Sigma_c$ except for the additional three-body $\bar D\Lambda_c\pi$ decay of $\bar D\Sigma_c^*$ molecule. $\bar D^*\Lambda_c$ is still the largest decay channel of $S$-wave $\bar D\Sigma_c^*$ molecule. The difference of the decay patterns between two form factors $f_1$ and $f_2$ in $\bar D \Sigma_c^*$ and $\bar D^*\Sigma_c^*$ sectors is similar with $\bar D \Sigma_c$ and $\bar D^*\Sigma_c$ molecules. The non-relativistic form factor $f_2$ brings a larger $D$ and $D^*$ meson exchanged partial widths. In particular, for the $1/2^-$-$\bar D^*\Sigma_c^*$ molecule, a huge enhancement for the $D^*$ exchanged precesses in $J/\psi p$, $\rho N$, $\omega p$, $\chi_{c0} p$ channels and the $D$ exchanged precesses in $\pi p$, $\eta_c p$ channels is generated by $f_2$. And there are some intriguing results for three $\bar D^*\Sigma_c^*$ molecules. Among of them, the $1/2^-$-$\bar D^*\Sigma_c^*$ molecule has the strongest couplings to $\bar D\Lambda_c$, $\bar D\Sigma_c$, $\bar D^*\Sigma_c$ channels and the relative ratio is around $1:1:1$ while $3/2^-$-$\bar D^*\Sigma_c^*$ is strongly coupled to the $\bar D\Sigma_c^*$ channel. In addition, the relative ratio between $\bar D^*\Lambda_c$ and $\bar D\Lambda_c$ is also different for these three $S$-wave $\bar D^*\Sigma_c^*$ molecule, $\Gamma_{\bar D^*\Lambda_c}/\Gamma_{\bar D\Lambda_c}$ is a bit less than 1 for the $1/2^-$ state, $\Gamma_{\bar D^*\Lambda_c}/\Gamma_{\bar D\Lambda_c}=3$ for the $5/2^-$ and it is around 50 for the $3/2^-$ molecule. The results obtained here can expand our understanding on the nature of pentaquark states in the hadronic molecule scenarios and can serve as the theoretical references for testing the molecule interpretations in the future experiments.
%---------
%---------
%-----------
\begin{table}[htpb]
	\centering
	\caption{\label{table:widths3}The partial decay widths of $P_c(4376)$ as $S$-wave $\bar{D}\Sigma_c^*$ molecule and $P_c(4500)$, $P_c(4511)$ and $P_c(4523)$ as the $\bar{D}^*\Sigma_c^*$ molecule with different spin parity which are four spin partners of observed $P_c$ molecules within the HQSS framework. The form factor set $(f_1,f_3)$ is used and $\Lambda_0=1.0\ \mathrm{GeV}$, $\Lambda_1=0.6\ \mathrm{GeV}$. The notation is same with Table.~\ref{table:widths1}.}
	\begin{tabular}{l|*{4}{c}}
		\Xhline{1pt}
		\multirow{3}*{Mode} & \multicolumn{4}{c}{Widths ($\mathrm{MeV}$) with $(f_1,f_3)$} \\
		\Xcline{2-5}{0.4pt}
		& \multicolumn{1}{c}{$\bar D \Sigma_c^*$} & \multicolumn{3}{c}{$\bar D^*\Sigma_c^*$} \\
		\Xcline{2-2}{0.4pt}\Xcline{3-5}{0.4pt}
		& \multicolumn{1}{c}{$P_c(4376)$}& \multicolumn{1}{c}{$P_c(4500)$} & \multicolumn{1}{c}{$P_c(4511)$} & \multicolumn{1}{c}{$P_c(4523)$} \\
		\Xcline{2-2}{0.4pt}\Xcline{3-3}{0.4pt}\Xcline{4-4}{0.4pt}\Xcline{5-5}{0.4pt}
		& \multicolumn{1}{c}{${\frac32}^-$}& \multicolumn{1}{c}{${\frac12}^-$} & \multicolumn{1}{c}{${\frac32}^-$} & \multicolumn{1}{c}{${\frac52}^-$} \\
		\Xhline{0.8pt}
		$\bar D^*\Lambda_c$ 	 &12.4 &7.1 &17.0 &4.5 \\
		$J/\psi p$ 		 	     &0.01 &0.006 &0.02 &0.006 \\
		$\bar D\Lambda_c$  	     &$9\time10^{-5}$ &10.0 &0.3 &1.5 \\
		$\pi N$ 			 	 &$2\time10^{-4}$ &0.003 &$1\time10^{-4}$ &$3\time10^{-4}$ \\
		$\chi_{c0}p$ 		 	 &0.003 &0.01 &0.002 &$6\time10^{-7}$ \\
		$\eta_c p$ 		 	     &0.001 &0.01 &$6\time10^{-4}$ &$8\time10^{-4}$ \\
		$\rho N$ 			  	 &$5\time10^{-4}$ &0.001 &0.01 &$8\time10^{-5}$ \\
		$\omega p$ 		 	     &0.002 &0.004 &0.005 &$3\time10^{-4}$ \\
		$\bar D\Sigma_c$ 	  	 &$5\time10^{-4}$ &10.6 &0.2 &1.3 \\
		$\bar D\Sigma^*_c$ 	  	 &- &1.0 &33.8 &6.2 \\
		$\bar D^*\Sigma_c$ 	  	 &- &10.6 &0.07 &1.2 \\
		$\bar D\Lambda_c \pi$ 	 &5.0 &- &- &- \\
		$\bar D^*\Lambda_c \pi$  &- &4.0 &7.7 &7.8 \\
		\Xhline{0.8pt}
		Total 				 	 &17.5 &43.3 &59.1 &22.5 \\
		\Xhline{1pt}
	\end{tabular}
\end{table}
%-----------
%-----------
\begin{table}[htpb]
	\centering
	\caption{\label{table:widths4}The numerical results for the form factor set $(f_2,f_3)$. The notation is same with Table.~\ref{table:widths3}.}
	\begin{tabular}{l|*{4}{c}}
		\Xhline{1pt}
		\multirow{3}*{Mode} & \multicolumn{4}{c}{Widths ($\mathrm{MeV}$) with $(f_2,f_3)$} \\
		\Xcline{2-5}{0.4pt}
		& \multicolumn{1}{c}{$\bar D \Sigma_c^*$} & \multicolumn{3}{c}{$\bar D^*\Sigma_c^*$} \\
		\Xcline{2-2}{0.4pt}\Xcline{3-5}{0.4pt}
		& \multicolumn{1}{c}{$P_c(4376)$}& \multicolumn{1}{c}{$P_c(4500)$} & \multicolumn{1}{c}{$P_c(4511)$} & \multicolumn{1}{c}{$P_c(4523)$} \\
		\Xcline{2-2}{0.4pt}\Xcline{3-3}{0.4pt}\Xcline{4-4}{0.4pt}\Xcline{5-5}{0.4pt}
		& \multicolumn{1}{c}{${\frac32}^-$}& \multicolumn{1}{c}{${\frac12}^-$} & \multicolumn{1}{c}{${\frac32}^-$} & \multicolumn{1}{c}{${\frac52}^-$} \\
		\Xhline{0.8pt}
		$\bar D^*\Lambda_c$ 	 &21.6 &6.4 &16.7 &3.1 \\
		$J/\psi p$ 		 	     &0.7 &36.7 &4.4 &0.2 \\
		$\bar D\Lambda_c$  	     &$3\time10^{-5}$ &2.0 &0.09 &0.7 \\
		$\pi N$ 			 	 &0.6 &49.9 &6.0 &0.5 \\
		$\chi_{c0}p$ 		 	 &0.1 &4.7 &0.5 &$8\time10^{-6}$ \\
		$\eta_c p$ 		 	     &$9\time10^{-4}$ &13.5 &0.1 &0.04 \\
		$\rho N$ 			  	 &0.2 &11.6 &0.6 &0.1 \\
		$\omega p$ 		 	     &0.8 &44.0 &2.3 &0.4 \\
		$\bar D\Sigma_c$ 	  	 &$2\time10^{-4}$ &6.7 &0.2 &1.0 \\
		$\bar D\Sigma^*_c$ 	  	 &- &1.2 &35.0 &4.1 \\
		$\bar D^*\Sigma_c$ 	  	 &- &13.6 &0.08 &0.7 \\
		$\bar D\Lambda_c \pi$ 	 &5.0 &- &- &- \\
		$\bar D^*\Lambda_c \pi$  &- &4.0 &7.7 &7.8 \\
		\Xhline{0.8pt}
		Total 				 	 &29.0 &194.5 &73.7 &18.7 \\
		\Xhline{1pt}
	\end{tabular}
\end{table}
%-----------

\section{Summary}\label{sec:summary}
A more precise spectrum of pentaquark-like states in the process of $\Lambda_b\to J/\psi p K$ was reported recently by the LHCb collaboration. As previous discovery of $P_c(4380)$ and $P_c(4450)$, the newly observed $P_c(4312)$, $P_c(4440)$ and $P_c(4457)$ have sparked a heated discussion. Inspired by the closeness of $P_c(4312)$ to the threshold of $\bar D\Sigma_c$ and $P_c(4440)$, $P_c(4457)$ states to the $\bar D^*\Sigma_c$, the natural hadronic molecular interpretation has been suggested in many theoretical works for these states. In analogy to our previous work on the $P_c(4380)$ and $P_c(4450)$, we investigate the strong decays of these newly observed $P_c$ states in the molecule scenarios. With the effective Lagrangian approach, the partial decay widths of $P_c$ states to all possible allowed channels are presented. It is found that the measured widths of $P_c(4312)$, $P_c(4440)$ and $P_c(4457)$ can be reproduced well respectively in the $1/2^-$-$\bar D\Sigma_c$, $1/2^-$-$\bar D^*\Sigma_c$ and $3/2^-$-$\bar D^*\Sigma_c$ molecule pictures. And the $3/2^-$-$\bar D^*\Sigma_c$ and $1/2^-$-$\bar D^*\Sigma_c$ molecule assignments for the $P_c(4440)$ and $P_c(4457)$ can not be ruled out at present. The novel difference on the decay patterns between the spin parity $1/2^-$ and $3/2^-$ $\bar D^*\Sigma_c$ molecule for both $P_c(4440)$ and $P_c(4457)$, such as $\Gamma_{\bar D\Sigma_c}/\Gamma_{\bar D\Sigma_c^*}$ and $\Gamma_{J/\psi p}/\Gamma_{\eta_c p}$, can be used to distinguish the quantum numbers in the future experiments. In addition, four possible heavy quark multiplets are also considered in our calculations. With the same cutoffs, their partial decay widths are presented. Albeit with large uncertainty, the findings here discussed can be considered as the direct consequences of the hadronic molecular assignments and can be tested by the further experimental investigation in future. It will improve our understanding on the inner structure of these pentaquark-like states.

\bigskip
\noindent
\begin{center}
	{\bf ACKNOWLEDGEMENTS}\\
	
\end{center}
We thank Cheng-Jian Xiao, Yin Huang, Feng-Kun Guo and Jia-Jun Wu for helpful discussions. This project is supported by NSFC under Grant No.~11621131001 (CRC110 cofunded by DFG and NSFC) and Grant No.~11747601, and the Chinese Academy of Sciences (CAS) under Grant No. XDPB09.

\bibliography{pc-refs}

\begin{thebibliography}{70}%
\makeatletter
\providecommand \@ifxundefined [1]{%
 \@ifx{#1\undefined}
}%
\providecommand \@ifnum [1]{%
 \ifnum #1\expandafter \@firstoftwo
 \else \expandafter \@secondoftwo
 \fi
}%
\providecommand \@ifx [1]{%
 \ifx #1\expandafter \@firstoftwo
 \else \expandafter \@secondoftwo
 \fi
}%
\providecommand \natexlab [1]{#1}%
\providecommand \enquote  [1]{``#1''}%
\providecommand \bibnamefont  [1]{#1}%
\providecommand \bibfnamefont [1]{#1}%
\providecommand \citenamefont [1]{#1}%
\providecommand \href@noop [0]{\@secondoftwo}%
\providecommand \href [0]{\begingroup \@sanitize@url \@href}%
\providecommand \@href[1]{\@@startlink{#1}\@@href}%
\providecommand \@@href[1]{\endgroup#1\@@endlink}%
\providecommand \@sanitize@url [0]{\catcode `\\12\catcode `\$12\catcode
  `\&12\catcode `\#12\catcode `\^12\catcode `\_12\catcode `\%12\relax}%
\providecommand \@@startlink[1]{}%
\providecommand \@@endlink[0]{}%
\providecommand \url  [0]{\begingroup\@sanitize@url \@url }%
\providecommand \@url [1]{\endgroup\@href {#1}{\urlprefix }}%
\providecommand \urlprefix  [0]{URL }%
\providecommand \Eprint [0]{\href }%
\providecommand \doibase [0]{http://dx.doi.org/}%
\providecommand \selectlanguage [0]{\@gobble}%
\providecommand \bibinfo  [0]{\@secondoftwo}%
\providecommand \bibfield  [0]{\@secondoftwo}%
\providecommand \translation [1]{[#1]}%
\providecommand \BibitemOpen [0]{}%
\providecommand \bibitemStop [0]{}%
\providecommand \bibitemNoStop [0]{.\EOS\space}%
\providecommand \EOS [0]{\spacefactor3000\relax}%
\providecommand \BibitemShut  [1]{\csname bibitem#1\endcsname}%
\let\auto@bib@innerbib\@empty
%</preamble>
\bibitem [{\citenamefont {Aaij}\ \emph {et~al.}(2019)\citenamefont {Aaij} \emph
  {et~al.}}]{Aaij:2019vzc}%
  \BibitemOpen
  \bibfield  {author} {\bibinfo {author} {\bibfnamefont {R.}~\bibnamefont
  {Aaij}} \emph {et~al.} (\bibinfo {collaboration} {LHCb}),\ }\href {\doibase
  10.1103/PhysRevLett.122.222001} {\bibfield  {journal} {\bibinfo  {journal}
  {Phys. Rev. Lett.}\ }\textbf {\bibinfo {volume} {122}},\ \bibinfo {pages}
  {222001} (\bibinfo {year} {2019})},\ \Eprint
  {http://arxiv.org/abs/1904.03947} {arXiv:1904.03947 [hep-ex]} \BibitemShut
  {NoStop}%
%%CITATION = ARXIV:1904.03947;%%
\bibitem [{\citenamefont {Lin}\ \emph {et~al.}(2017)\citenamefont {Lin},
  \citenamefont {Shen}, \citenamefont {Guo},\ and\ \citenamefont
  {Zou}}]{Lin:2017mtz}%
  \BibitemOpen
  \bibfield  {author} {\bibinfo {author} {\bibfnamefont {Y.-H.}\ \bibnamefont
  {Lin}}, \bibinfo {author} {\bibfnamefont {C.-W.}\ \bibnamefont {Shen}},
  \bibinfo {author} {\bibfnamefont {F.-K.}\ \bibnamefont {Guo}}, \ and\
  \bibinfo {author} {\bibfnamefont {B.-S.}\ \bibnamefont {Zou}},\ }\href
  {\doibase 10.1103/PhysRevD.95.114017} {\bibfield  {journal} {\bibinfo
  {journal} {Phys. Rev.}\ }\textbf {\bibinfo {volume} {D95}},\ \bibinfo {pages}
  {114017} (\bibinfo {year} {2017})},\ \Eprint
  {http://arxiv.org/abs/1703.01045} {arXiv:1703.01045 [hep-ph]} \BibitemShut
  {NoStop}%
%%CITATION = ARXIV:1703.01045;%%
\bibitem [{\citenamefont {Aaij}\ \emph {et~al.}(2015)\citenamefont {Aaij} \emph
  {et~al.}}]{Aaij:2015tga}%
  \BibitemOpen
  \bibfield  {author} {\bibinfo {author} {\bibfnamefont {R.}~\bibnamefont
  {Aaij}} \emph {et~al.} (\bibinfo {collaboration} {LHCb}),\ }\href {\doibase
  10.1103/PhysRevLett.115.072001} {\bibfield  {journal} {\bibinfo  {journal}
  {Phys. Rev. Lett.}\ }\textbf {\bibinfo {volume} {115}},\ \bibinfo {pages}
  {072001} (\bibinfo {year} {2015})},\ \Eprint
  {http://arxiv.org/abs/1507.03414} {arXiv:1507.03414 [hep-ex]} \BibitemShut
  {NoStop}%
%%CITATION = ARXIV:1507.03414;%%
\bibitem [{\citenamefont {Wu}\ \emph {et~al.}(2010)\citenamefont {Wu},
  \citenamefont {Molina}, \citenamefont {Oset},\ and\ \citenamefont
  {Zou}}]{Wu:2010jy}%
  \BibitemOpen
  \bibfield  {author} {\bibinfo {author} {\bibfnamefont {J.-J.}\ \bibnamefont
  {Wu}}, \bibinfo {author} {\bibfnamefont {R.}~\bibnamefont {Molina}}, \bibinfo
  {author} {\bibfnamefont {E.}~\bibnamefont {Oset}}, \ and\ \bibinfo {author}
  {\bibfnamefont {B.~S.}\ \bibnamefont {Zou}},\ }\href {\doibase
  10.1103/PhysRevLett.105.232001} {\bibfield  {journal} {\bibinfo  {journal}
  {Phys. Rev. Lett.}\ }\textbf {\bibinfo {volume} {105}},\ \bibinfo {pages}
  {232001} (\bibinfo {year} {2010})},\ \Eprint {http://arxiv.org/abs/1007.0573}
  {arXiv:1007.0573 [nucl-th]} \BibitemShut {NoStop}%
%%CITATION = ARXIV:1007.0573;%%
\bibitem [{\citenamefont {Wu}\ \emph {et~al.}(2011)\citenamefont {Wu},
  \citenamefont {Molina}, \citenamefont {Oset},\ and\ \citenamefont
  {Zou}}]{Wu:2010vk}%
  \BibitemOpen
  \bibfield  {author} {\bibinfo {author} {\bibfnamefont {J.-J.}\ \bibnamefont
  {Wu}}, \bibinfo {author} {\bibfnamefont {R.}~\bibnamefont {Molina}}, \bibinfo
  {author} {\bibfnamefont {E.}~\bibnamefont {Oset}}, \ and\ \bibinfo {author}
  {\bibfnamefont {B.~S.}\ \bibnamefont {Zou}},\ }\href {\doibase
  10.1103/PhysRevC.84.015202} {\bibfield  {journal} {\bibinfo  {journal} {Phys.
  Rev.}\ }\textbf {\bibinfo {volume} {C84}},\ \bibinfo {pages} {015202}
  (\bibinfo {year} {2011})},\ \Eprint {http://arxiv.org/abs/1011.2399}
  {arXiv:1011.2399 [nucl-th]} \BibitemShut {NoStop}%
%%CITATION = ARXIV:1011.2399;%%
\bibitem [{\citenamefont {Wang}\ \emph {et~al.}(2011)\citenamefont {Wang},
  \citenamefont {Huang}, \citenamefont {Zhang},\ and\ \citenamefont
  {Zou}}]{Wang:2011rga}%
  \BibitemOpen
  \bibfield  {author} {\bibinfo {author} {\bibfnamefont {W.~L.}\ \bibnamefont
  {Wang}}, \bibinfo {author} {\bibfnamefont {F.}~\bibnamefont {Huang}},
  \bibinfo {author} {\bibfnamefont {Z.~Y.}\ \bibnamefont {Zhang}}, \ and\
  \bibinfo {author} {\bibfnamefont {B.~S.}\ \bibnamefont {Zou}},\ }\href
  {\doibase 10.1103/PhysRevC.84.015203} {\bibfield  {journal} {\bibinfo
  {journal} {Phys. Rev.}\ }\textbf {\bibinfo {volume} {C84}},\ \bibinfo {pages}
  {015203} (\bibinfo {year} {2011})},\ \Eprint {http://arxiv.org/abs/1101.0453}
  {arXiv:1101.0453 [nucl-th]} \BibitemShut {NoStop}%
%%CITATION = ARXIV:1101.0453;%%
\bibitem [{\citenamefont {Yang}\ \emph {et~al.}(2012)\citenamefont {Yang},
  \citenamefont {Sun}, \citenamefont {He}, \citenamefont {Liu},\ and\
  \citenamefont {Zhu}}]{Yang:2011wz}%
  \BibitemOpen
  \bibfield  {author} {\bibinfo {author} {\bibfnamefont {Z.-C.}\ \bibnamefont
  {Yang}}, \bibinfo {author} {\bibfnamefont {Z.-F.}\ \bibnamefont {Sun}},
  \bibinfo {author} {\bibfnamefont {J.}~\bibnamefont {He}}, \bibinfo {author}
  {\bibfnamefont {X.}~\bibnamefont {Liu}}, \ and\ \bibinfo {author}
  {\bibfnamefont {S.-L.}\ \bibnamefont {Zhu}},\ }\href {\doibase
  10.1088/1674-1137/36/1/002, 10.1088/1674-1137/36/3/006} {\bibfield  {journal}
  {\bibinfo  {journal} {Chin. Phys.}\ }\textbf {\bibinfo {volume} {C36}},\
  \bibinfo {pages} {6} (\bibinfo {year} {2012})},\ \Eprint
  {http://arxiv.org/abs/1105.2901} {arXiv:1105.2901 [hep-ph]} \BibitemShut
  {NoStop}%
%%CITATION = ARXIV:1105.2901;%%
\bibitem [{\citenamefont {Wu}\ \emph {et~al.}(2012)\citenamefont {Wu},
  \citenamefont {Lee},\ and\ \citenamefont {Zou}}]{Wu:2012md}%
  \BibitemOpen
  \bibfield  {author} {\bibinfo {author} {\bibfnamefont {J.-J.}\ \bibnamefont
  {Wu}}, \bibinfo {author} {\bibfnamefont {T.~S.~H.}\ \bibnamefont {Lee}}, \
  and\ \bibinfo {author} {\bibfnamefont {B.~S.}\ \bibnamefont {Zou}},\ }\href
  {\doibase 10.1103/PhysRevC.85.044002} {\bibfield  {journal} {\bibinfo
  {journal} {Phys. Rev.}\ }\textbf {\bibinfo {volume} {C85}},\ \bibinfo {pages}
  {044002} (\bibinfo {year} {2012})},\ \Eprint {http://arxiv.org/abs/1202.1036}
  {arXiv:1202.1036 [nucl-th]} \BibitemShut {NoStop}%
%%CITATION = ARXIV:1202.1036;%%
\bibitem [{\citenamefont {Yuan}\ \emph {et~al.}(2012)\citenamefont {Yuan},
  \citenamefont {Wei}, \citenamefont {He}, \citenamefont {Xu},\ and\
  \citenamefont {Zou}}]{Yuan:2012wz}%
  \BibitemOpen
  \bibfield  {author} {\bibinfo {author} {\bibfnamefont {S.~G.}\ \bibnamefont
  {Yuan}}, \bibinfo {author} {\bibfnamefont {K.~W.}\ \bibnamefont {Wei}},
  \bibinfo {author} {\bibfnamefont {J.}~\bibnamefont {He}}, \bibinfo {author}
  {\bibfnamefont {H.~S.}\ \bibnamefont {Xu}}, \ and\ \bibinfo {author}
  {\bibfnamefont {B.~S.}\ \bibnamefont {Zou}},\ }\href {\doibase
  10.1140/epja/i2012-12061-2} {\bibfield  {journal} {\bibinfo  {journal} {Eur.
  Phys. J.}\ }\textbf {\bibinfo {volume} {A48}},\ \bibinfo {pages} {61}
  (\bibinfo {year} {2012})},\ \Eprint {http://arxiv.org/abs/1201.0807}
  {arXiv:1201.0807 [nucl-th]} \BibitemShut {NoStop}%
%%CITATION = ARXIV:1201.0807;%%
\bibitem [{\citenamefont {Xiao}\ \emph {et~al.}(2013)\citenamefont {Xiao},
  \citenamefont {Nieves},\ and\ \citenamefont {Oset}}]{Xiao:2013yca}%
  \BibitemOpen
  \bibfield  {author} {\bibinfo {author} {\bibfnamefont {C.~W.}\ \bibnamefont
  {Xiao}}, \bibinfo {author} {\bibfnamefont {J.}~\bibnamefont {Nieves}}, \ and\
  \bibinfo {author} {\bibfnamefont {E.}~\bibnamefont {Oset}},\ }\href {\doibase
  10.1103/PhysRevD.88.056012} {\bibfield  {journal} {\bibinfo  {journal} {Phys.
  Rev.}\ }\textbf {\bibinfo {volume} {D88}},\ \bibinfo {pages} {056012}
  (\bibinfo {year} {2013})},\ \Eprint {http://arxiv.org/abs/1304.5368}
  {arXiv:1304.5368 [hep-ph]} \BibitemShut {NoStop}%
%%CITATION = ARXIV:1304.5368;%%
\bibitem [{\citenamefont {Jaffe}\ and\ \citenamefont
  {Wilczek}(2003)}]{Jaffe:2003sg}%
  \BibitemOpen
  \bibfield  {author} {\bibinfo {author} {\bibfnamefont {R.~L.}\ \bibnamefont
  {Jaffe}}\ and\ \bibinfo {author} {\bibfnamefont {F.}~\bibnamefont
  {Wilczek}},\ }\href {\doibase 10.1103/PhysRevLett.91.232003} {\bibfield
  {journal} {\bibinfo  {journal} {Phys. Rev. Lett.}\ }\textbf {\bibinfo
  {volume} {91}},\ \bibinfo {pages} {232003} (\bibinfo {year} {2003})},\
  \Eprint {http://arxiv.org/abs/hep-ph/0307341} {arXiv:hep-ph/0307341 [hep-ph]}
  \BibitemShut {NoStop}%
%%CITATION = HEP-PH/0307341;%%
\bibitem [{\citenamefont {Ali}\ \emph {et~al.}(2016)\citenamefont {Ali},
  \citenamefont {Ahmed}, \citenamefont {Aslam},\ and\ \citenamefont
  {Rehman}}]{Ali:2016dkf}%
  \BibitemOpen
  \bibfield  {author} {\bibinfo {author} {\bibfnamefont {A.}~\bibnamefont
  {Ali}}, \bibinfo {author} {\bibfnamefont {I.}~\bibnamefont {Ahmed}}, \bibinfo
  {author} {\bibfnamefont {M.~J.}\ \bibnamefont {Aslam}}, \ and\ \bibinfo
  {author} {\bibfnamefont {A.}~\bibnamefont {Rehman}},\ }\href {\doibase
  10.1103/PhysRevD.94.054001} {\bibfield  {journal} {\bibinfo  {journal} {Phys.
  Rev.}\ }\textbf {\bibinfo {volume} {D94}},\ \bibinfo {pages} {054001}
  (\bibinfo {year} {2016})},\ \Eprint {http://arxiv.org/abs/1607.00987}
  {arXiv:1607.00987 [hep-ph]} \BibitemShut {NoStop}%
%%CITATION = ARXIV:1607.00987;%%
\bibitem [{\citenamefont {Maiani}\ \emph {et~al.}(2015)\citenamefont {Maiani},
  \citenamefont {Polosa},\ and\ \citenamefont {Riquer}}]{Maiani:2015vwa}%
  \BibitemOpen
  \bibfield  {author} {\bibinfo {author} {\bibfnamefont {L.}~\bibnamefont
  {Maiani}}, \bibinfo {author} {\bibfnamefont {A.~D.}\ \bibnamefont {Polosa}},
  \ and\ \bibinfo {author} {\bibfnamefont {V.}~\bibnamefont {Riquer}},\ }\href
  {\doibase 10.1016/j.physletb.2015.08.008} {\bibfield  {journal} {\bibinfo
  {journal} {Phys. Lett.}\ }\textbf {\bibinfo {volume} {B749}},\ \bibinfo
  {pages} {289} (\bibinfo {year} {2015})},\ \Eprint
  {http://arxiv.org/abs/1507.04980} {arXiv:1507.04980 [hep-ph]} \BibitemShut
  {NoStop}%
%%CITATION = ARXIV:1507.04980;%%
\bibitem [{\citenamefont {Li}\ \emph {et~al.}(2015)\citenamefont {Li},
  \citenamefont {He},\ and\ \citenamefont {He}}]{Li:2015gta}%
  \BibitemOpen
  \bibfield  {author} {\bibinfo {author} {\bibfnamefont {G.-N.}\ \bibnamefont
  {Li}}, \bibinfo {author} {\bibfnamefont {X.-G.}\ \bibnamefont {He}}, \ and\
  \bibinfo {author} {\bibfnamefont {M.}~\bibnamefont {He}},\ }\href {\doibase
  10.1007/JHEP12(2015)128} {\bibfield  {journal} {\bibinfo  {journal} {JHEP}\
  }\textbf {\bibinfo {volume} {12}},\ \bibinfo {pages} {128} (\bibinfo {year}
  {2015})},\ \Eprint {http://arxiv.org/abs/1507.08252} {arXiv:1507.08252
  [hep-ph]} \BibitemShut {NoStop}%
%%CITATION = ARXIV:1507.08252;%%
\bibitem [{\citenamefont {Wang}(2016)}]{Wang:2015epa}%
  \BibitemOpen
  \bibfield  {author} {\bibinfo {author} {\bibfnamefont {Z.-G.}\ \bibnamefont
  {Wang}},\ }\href {\doibase 10.1140/epjc/s10052-016-3920-4} {\bibfield
  {journal} {\bibinfo  {journal} {Eur. Phys. J.}\ }\textbf {\bibinfo {volume}
  {C76}},\ \bibinfo {pages} {70} (\bibinfo {year} {2016})},\ \Eprint
  {http://arxiv.org/abs/1508.01468} {arXiv:1508.01468 [hep-ph]} \BibitemShut
  {NoStop}%
%%CITATION = ARXIV:1508.01468;%%
\bibitem [{\citenamefont {Weng}\ \emph {et~al.}(2019)\citenamefont {Weng},
  \citenamefont {Chen}, \citenamefont {Deng},\ and\ \citenamefont
  {Zhu}}]{Weng:2019ynv}%
  \BibitemOpen
  \bibfield  {author} {\bibinfo {author} {\bibfnamefont {X.-Z.}\ \bibnamefont
  {Weng}}, \bibinfo {author} {\bibfnamefont {X.-L.}\ \bibnamefont {Chen}},
  \bibinfo {author} {\bibfnamefont {W.-Z.}\ \bibnamefont {Deng}}, \ and\
  \bibinfo {author} {\bibfnamefont {S.-L.}\ \bibnamefont {Zhu}},\ }\href@noop
  {} {\  (\bibinfo {year} {2019})},\ \Eprint {http://arxiv.org/abs/1904.09891}
  {arXiv:1904.09891 [hep-ph]} \BibitemShut {NoStop}%
%%CITATION = ARXIV:1904.09891;%%
\bibitem [{\citenamefont {Stancu}(2019)}]{Stancu:2019qga}%
  \BibitemOpen
  \bibfield  {author} {\bibinfo {author} {\bibfnamefont {F.}~\bibnamefont
  {Stancu}},\ }\href@noop {} {\  (\bibinfo {year} {2019})},\ \Eprint
  {http://arxiv.org/abs/1902.07101} {arXiv:1902.07101 [hep-ph]} \BibitemShut
  {NoStop}%
%%CITATION = ARXIV:1902.07101;%%
\bibitem [{\citenamefont {Giannuzzi}(2019)}]{Giannuzzi:2019esi}%
  \BibitemOpen
  \bibfield  {author} {\bibinfo {author} {\bibfnamefont {F.}~\bibnamefont
  {Giannuzzi}},\ }\href {\doibase 10.1103/PhysRevD.99.094006} {\bibfield
  {journal} {\bibinfo  {journal} {Phys. Rev.}\ }\textbf {\bibinfo {volume}
  {D99}},\ \bibinfo {pages} {094006} (\bibinfo {year} {2019})},\ \Eprint
  {http://arxiv.org/abs/1903.04430} {arXiv:1903.04430 [hep-ph]} \BibitemShut
  {NoStop}%
%%CITATION = ARXIV:1903.04430;%%
\bibitem [{\citenamefont {Zhu}\ \emph {et~al.}(2019)\citenamefont {Zhu},
  \citenamefont {Liu}, \citenamefont {Huang},\ and\ \citenamefont
  {Qiao}}]{Zhu:2019iwm}%
  \BibitemOpen
  \bibfield  {author} {\bibinfo {author} {\bibfnamefont {R.}~\bibnamefont
  {Zhu}}, \bibinfo {author} {\bibfnamefont {X.}~\bibnamefont {Liu}}, \bibinfo
  {author} {\bibfnamefont {H.}~\bibnamefont {Huang}}, \ and\ \bibinfo {author}
  {\bibfnamefont {C.-F.}\ \bibnamefont {Qiao}},\ }\href@noop {} {\  (\bibinfo
  {year} {2019})},\ \Eprint {http://arxiv.org/abs/1904.10285} {arXiv:1904.10285
  [hep-ph]} \BibitemShut {NoStop}%
%%CITATION = ARXIV:1904.10285;%%
\bibitem [{\citenamefont {An}\ \emph {et~al.}(2019)\citenamefont {An},
  \citenamefont {Zhou}, \citenamefont {Liu}, \citenamefont {Liu},\ and\
  \citenamefont {Liu}}]{An:2019idk}%
  \BibitemOpen
  \bibfield  {author} {\bibinfo {author} {\bibfnamefont {H.-T.}\ \bibnamefont
  {An}}, \bibinfo {author} {\bibfnamefont {Q.-S.}\ \bibnamefont {Zhou}},
  \bibinfo {author} {\bibfnamefont {Z.-W.}\ \bibnamefont {Liu}}, \bibinfo
  {author} {\bibfnamefont {Y.-R.}\ \bibnamefont {Liu}}, \ and\ \bibinfo
  {author} {\bibfnamefont {X.}~\bibnamefont {Liu}},\ }\href@noop {} {\
  (\bibinfo {year} {2019})},\ \Eprint {http://arxiv.org/abs/1905.07858}
  {arXiv:1905.07858 [hep-ph]} \BibitemShut {NoStop}%
%%CITATION = ARXIV:1905.07858;%%
\bibitem [{\citenamefont {Kubarovsky}\ and\ \citenamefont
  {Voloshin}(2015)}]{Kubarovsky:2015aaa}%
  \BibitemOpen
  \bibfield  {author} {\bibinfo {author} {\bibfnamefont {V.}~\bibnamefont
  {Kubarovsky}}\ and\ \bibinfo {author} {\bibfnamefont {M.~B.}\ \bibnamefont
  {Voloshin}},\ }\href {\doibase 10.1103/PhysRevD.92.031502} {\bibfield
  {journal} {\bibinfo  {journal} {Phys. Rev.}\ }\textbf {\bibinfo {volume}
  {D92}},\ \bibinfo {pages} {031502} (\bibinfo {year} {2015})},\ \Eprint
  {http://arxiv.org/abs/1508.00888} {arXiv:1508.00888 [hep-ph]} \BibitemShut
  {NoStop}%
%%CITATION = ARXIV:1508.00888;%%
\bibitem [{\citenamefont {Eides}\ \emph {et~al.}(2019)\citenamefont {Eides},
  \citenamefont {Petrov},\ and\ \citenamefont {Polyakov}}]{Eides:2019tgv}%
  \BibitemOpen
  \bibfield  {author} {\bibinfo {author} {\bibfnamefont {M.~I.}\ \bibnamefont
  {Eides}}, \bibinfo {author} {\bibfnamefont {V.~Y.}\ \bibnamefont {Petrov}}, \
  and\ \bibinfo {author} {\bibfnamefont {M.~V.}\ \bibnamefont {Polyakov}},\
  }\href@noop {} {\  (\bibinfo {year} {2019})},\ \Eprint
  {http://arxiv.org/abs/1904.11616} {arXiv:1904.11616 [hep-ph]} \BibitemShut
  {NoStop}%
%%CITATION = ARXIV:1904.11616;%%
\bibitem [{\citenamefont {Guo}\ \emph {et~al.}(2015)\citenamefont {Guo},
  \citenamefont {Meißner}, \citenamefont {Wang},\ and\ \citenamefont
  {Yang}}]{Guo:2015umn}%
  \BibitemOpen
  \bibfield  {author} {\bibinfo {author} {\bibfnamefont {F.-K.}\ \bibnamefont
  {Guo}}, \bibinfo {author} {\bibfnamefont {U.-G.}\ \bibnamefont {Meißner}},
  \bibinfo {author} {\bibfnamefont {W.}~\bibnamefont {Wang}}, \ and\ \bibinfo
  {author} {\bibfnamefont {Z.}~\bibnamefont {Yang}},\ }\href {\doibase
  10.1103/PhysRevD.92.071502} {\bibfield  {journal} {\bibinfo  {journal} {Phys.
  Rev.}\ }\textbf {\bibinfo {volume} {D92}},\ \bibinfo {pages} {071502}
  (\bibinfo {year} {2015})},\ \Eprint {http://arxiv.org/abs/1507.04950}
  {arXiv:1507.04950 [hep-ph]} \BibitemShut {NoStop}%
%%CITATION = ARXIV:1507.04950;%%
\bibitem [{\citenamefont {Liu}\ \emph {et~al.}(2016)\citenamefont {Liu},
  \citenamefont {Wang},\ and\ \citenamefont {Zhao}}]{Liu:2015fea}%
  \BibitemOpen
  \bibfield  {author} {\bibinfo {author} {\bibfnamefont {X.-H.}\ \bibnamefont
  {Liu}}, \bibinfo {author} {\bibfnamefont {Q.}~\bibnamefont {Wang}}, \ and\
  \bibinfo {author} {\bibfnamefont {Q.}~\bibnamefont {Zhao}},\ }\href {\doibase
  10.1016/j.physletb.2016.03.089} {\bibfield  {journal} {\bibinfo  {journal}
  {Phys. Lett.}\ }\textbf {\bibinfo {volume} {B757}},\ \bibinfo {pages} {231}
  (\bibinfo {year} {2016})},\ \Eprint {http://arxiv.org/abs/1507.05359}
  {arXiv:1507.05359 [hep-ph]} \BibitemShut {NoStop}%
%%CITATION = ARXIV:1507.05359;%%
\bibitem [{\citenamefont {Guo}\ \emph {et~al.}(2016)\citenamefont {Guo},
  \citenamefont {Meißner}, \citenamefont {Nieves},\ and\ \citenamefont
  {Yang}}]{Guo:2016bkl}%
  \BibitemOpen
  \bibfield  {author} {\bibinfo {author} {\bibfnamefont {F.-K.}\ \bibnamefont
  {Guo}}, \bibinfo {author} {\bibfnamefont {U.~G.}\ \bibnamefont {Meißner}},
  \bibinfo {author} {\bibfnamefont {J.}~\bibnamefont {Nieves}}, \ and\ \bibinfo
  {author} {\bibfnamefont {Z.}~\bibnamefont {Yang}},\ }\href {\doibase
  10.1140/epja/i2016-16318-4} {\bibfield  {journal} {\bibinfo  {journal} {Eur.
  Phys. J.}\ }\textbf {\bibinfo {volume} {A52}},\ \bibinfo {pages} {318}
  (\bibinfo {year} {2016})},\ \Eprint {http://arxiv.org/abs/1605.05113}
  {arXiv:1605.05113 [hep-ph]} \BibitemShut {NoStop}%
%%CITATION = ARXIV:1605.05113;%%
\bibitem [{\citenamefont {Bayar}\ \emph {et~al.}(2016)\citenamefont {Bayar},
  \citenamefont {Aceti}, \citenamefont {Guo},\ and\ \citenamefont
  {Oset}}]{Bayar:2016ftu}%
  \BibitemOpen
  \bibfield  {author} {\bibinfo {author} {\bibfnamefont {M.}~\bibnamefont
  {Bayar}}, \bibinfo {author} {\bibfnamefont {F.}~\bibnamefont {Aceti}},
  \bibinfo {author} {\bibfnamefont {F.-K.}\ \bibnamefont {Guo}}, \ and\
  \bibinfo {author} {\bibfnamefont {E.}~\bibnamefont {Oset}},\ }\href {\doibase
  10.1103/PhysRevD.94.074039} {\bibfield  {journal} {\bibinfo  {journal} {Phys.
  Rev.}\ }\textbf {\bibinfo {volume} {D94}},\ \bibinfo {pages} {074039}
  (\bibinfo {year} {2016})},\ \Eprint {http://arxiv.org/abs/1609.04133}
  {arXiv:1609.04133 [hep-ph]} \BibitemShut {NoStop}%
%%CITATION = ARXIV:1609.04133;%%
\bibitem [{\citenamefont {Mironov}\ and\ \citenamefont
  {Morozov}(2015)}]{Mironov:2015ica}%
  \BibitemOpen
  \bibfield  {author} {\bibinfo {author} {\bibfnamefont {A.}~\bibnamefont
  {Mironov}}\ and\ \bibinfo {author} {\bibfnamefont {A.}~\bibnamefont
  {Morozov}},\ }\href {\doibase 10.7868/S0370274X15170038,
  10.1134/S0021364015170099} {\bibfield  {journal} {\bibinfo  {journal} {JETP
  Lett.}\ }\textbf {\bibinfo {volume} {102}},\ \bibinfo {pages} {271} (\bibinfo
  {year} {2015})},\ \bibinfo {note} {[Pisma Zh. Eksp. Teor.
  Fiz.102,no.5,302(2015)]},\ \Eprint {http://arxiv.org/abs/1507.04694}
  {arXiv:1507.04694 [hep-ph]} \BibitemShut {NoStop}%
%%CITATION = ARXIV:1507.04694;%%
\bibitem [{\citenamefont {Scoccola}\ \emph {et~al.}(2015)\citenamefont
  {Scoccola}, \citenamefont {Riska},\ and\ \citenamefont
  {Rho}}]{Scoccola:2015nia}%
  \BibitemOpen
  \bibfield  {author} {\bibinfo {author} {\bibfnamefont {N.~N.}\ \bibnamefont
  {Scoccola}}, \bibinfo {author} {\bibfnamefont {D.~O.}\ \bibnamefont {Riska}},
  \ and\ \bibinfo {author} {\bibfnamefont {M.}~\bibnamefont {Rho}},\ }\href
  {\doibase 10.1103/PhysRevD.92.051501} {\bibfield  {journal} {\bibinfo
  {journal} {Phys. Rev.}\ }\textbf {\bibinfo {volume} {D92}},\ \bibinfo {pages}
  {051501} (\bibinfo {year} {2015})},\ \Eprint
  {http://arxiv.org/abs/1508.01172} {arXiv:1508.01172 [hep-ph]} \BibitemShut
  {NoStop}%
%%CITATION = ARXIV:1508.01172;%%
\bibitem [{\citenamefont {Chen}\ \emph
  {et~al.}(2019{\natexlab{a}})\citenamefont {Chen}, \citenamefont {Chen},\ and\
  \citenamefont {Zhu}}]{Chen:2019bip}%
  \BibitemOpen
  \bibfield  {author} {\bibinfo {author} {\bibfnamefont {H.-X.}\ \bibnamefont
  {Chen}}, \bibinfo {author} {\bibfnamefont {W.}~\bibnamefont {Chen}}, \ and\
  \bibinfo {author} {\bibfnamefont {S.-L.}\ \bibnamefont {Zhu}},\ }\href@noop
  {} {\  (\bibinfo {year} {2019}{\natexlab{a}})},\ \Eprint
  {http://arxiv.org/abs/1903.11001} {arXiv:1903.11001 [hep-ph]} \BibitemShut
  {NoStop}%
%%CITATION = ARXIV:1903.11001;%%
\bibitem [{\citenamefont {Chen}\ \emph
  {et~al.}(2019{\natexlab{b}})\citenamefont {Chen}, \citenamefont {Sun},
  \citenamefont {Liu},\ and\ \citenamefont {Zhu}}]{Chen:2019asm}%
  \BibitemOpen
  \bibfield  {author} {\bibinfo {author} {\bibfnamefont {R.}~\bibnamefont
  {Chen}}, \bibinfo {author} {\bibfnamefont {Z.-F.}\ \bibnamefont {Sun}},
  \bibinfo {author} {\bibfnamefont {X.}~\bibnamefont {Liu}}, \ and\ \bibinfo
  {author} {\bibfnamefont {S.-L.}\ \bibnamefont {Zhu}},\ }\href@noop {} {\
  (\bibinfo {year} {2019}{\natexlab{b}})},\ \Eprint
  {http://arxiv.org/abs/1903.11013} {arXiv:1903.11013 [hep-ph]} \BibitemShut
  {NoStop}%
%%CITATION = ARXIV:1903.11013;%%
\bibitem [{\citenamefont {Guo}\ \emph {et~al.}(2019)\citenamefont {Guo},
  \citenamefont {Jing}, \citenamefont {Meißner},\ and\ \citenamefont
  {Sakai}}]{Guo:2019fdo}%
  \BibitemOpen
  \bibfield  {author} {\bibinfo {author} {\bibfnamefont {F.-K.}\ \bibnamefont
  {Guo}}, \bibinfo {author} {\bibfnamefont {H.-J.}\ \bibnamefont {Jing}},
  \bibinfo {author} {\bibfnamefont {U.-G.}\ \bibnamefont {Meißner}}, \ and\
  \bibinfo {author} {\bibfnamefont {S.}~\bibnamefont {Sakai}},\ }\href
  {\doibase 10.1103/PhysRevD.99.091501} {\bibfield  {journal} {\bibinfo
  {journal} {Phys. Rev.}\ }\textbf {\bibinfo {volume} {D99}},\ \bibinfo {pages}
  {091501} (\bibinfo {year} {2019})},\ \Eprint
  {http://arxiv.org/abs/1903.11503} {arXiv:1903.11503 [hep-ph]} \BibitemShut
  {NoStop}%
%%CITATION = ARXIV:1903.11503;%%
\bibitem [{\citenamefont {Liu}\ \emph {et~al.}(2019{\natexlab{a}})\citenamefont
  {Liu}, \citenamefont {Pan}, \citenamefont {Peng}, \citenamefont
  {Sánchez~Sánchez}, \citenamefont {Geng}, \citenamefont {Hosaka},\ and\
  \citenamefont {Pavon~Valderrama}}]{Liu:2019tjn}%
  \BibitemOpen
  \bibfield  {author} {\bibinfo {author} {\bibfnamefont {M.-Z.}\ \bibnamefont
  {Liu}}, \bibinfo {author} {\bibfnamefont {Y.-W.}\ \bibnamefont {Pan}},
  \bibinfo {author} {\bibfnamefont {F.-Z.}\ \bibnamefont {Peng}}, \bibinfo
  {author} {\bibfnamefont {M.}~\bibnamefont {Sánchez~Sánchez}}, \bibinfo
  {author} {\bibfnamefont {L.-S.}\ \bibnamefont {Geng}}, \bibinfo {author}
  {\bibfnamefont {A.}~\bibnamefont {Hosaka}}, \ and\ \bibinfo {author}
  {\bibfnamefont {M.}~\bibnamefont {Pavon~Valderrama}},\ }\href {\doibase
  10.1103/PhysRevLett.122.242001} {\bibfield  {journal} {\bibinfo  {journal}
  {Phys. Rev. Lett.}\ }\textbf {\bibinfo {volume} {122}},\ \bibinfo {pages}
  {242001} (\bibinfo {year} {2019}{\natexlab{a}})},\ \Eprint
  {http://arxiv.org/abs/1903.11560} {arXiv:1903.11560 [hep-ph]} \BibitemShut
  {NoStop}%
%%CITATION = ARXIV:1903.11560;%%
\bibitem [{\citenamefont {He}(2019)}]{He:2019ify}%
  \BibitemOpen
  \bibfield  {author} {\bibinfo {author} {\bibfnamefont {J.}~\bibnamefont
  {He}},\ }\href {\doibase 10.1140/epjc/s10052-019-6906-1} {\bibfield
  {journal} {\bibinfo  {journal} {Eur. Phys. J.}\ }\textbf {\bibinfo {volume}
  {C79}},\ \bibinfo {pages} {393} (\bibinfo {year} {2019})},\ \Eprint
  {http://arxiv.org/abs/1903.11872} {arXiv:1903.11872 [hep-ph]} \BibitemShut
  {NoStop}%
%%CITATION = ARXIV:1903.11872;%%
\bibitem [{\citenamefont {Liu}\ \emph {et~al.}(2019{\natexlab{b}})\citenamefont
  {Liu}, \citenamefont {Chen}, \citenamefont {Chen}, \citenamefont {Liu},\ and\
  \citenamefont {Zhu}}]{Liu:2019zoy}%
  \BibitemOpen
  \bibfield  {author} {\bibinfo {author} {\bibfnamefont {Y.-R.}\ \bibnamefont
  {Liu}}, \bibinfo {author} {\bibfnamefont {H.-X.}\ \bibnamefont {Chen}},
  \bibinfo {author} {\bibfnamefont {W.}~\bibnamefont {Chen}}, \bibinfo {author}
  {\bibfnamefont {X.}~\bibnamefont {Liu}}, \ and\ \bibinfo {author}
  {\bibfnamefont {S.-L.}\ \bibnamefont {Zhu}},\ }\href {\doibase
  10.1016/j.ppnp.2019.04.003} {\bibfield  {journal} {\bibinfo  {journal} {Prog.
  Part. Nucl. Phys.}\ }\textbf {\bibinfo {volume} {107}},\ \bibinfo {pages}
  {237} (\bibinfo {year} {2019}{\natexlab{b}})},\ \Eprint
  {http://arxiv.org/abs/1903.11976} {arXiv:1903.11976 [hep-ph]} \BibitemShut
  {NoStop}%
%%CITATION = ARXIV:1903.11976;%%
\bibitem [{\citenamefont {Huang}\ \emph {et~al.}(2019)\citenamefont {Huang},
  \citenamefont {He},\ and\ \citenamefont {Ping}}]{Huang:2019jlf}%
  \BibitemOpen
  \bibfield  {author} {\bibinfo {author} {\bibfnamefont {H.}~\bibnamefont
  {Huang}}, \bibinfo {author} {\bibfnamefont {J.}~\bibnamefont {He}}, \ and\
  \bibinfo {author} {\bibfnamefont {J.}~\bibnamefont {Ping}},\ }\href@noop {}
  {\  (\bibinfo {year} {2019})},\ \Eprint {http://arxiv.org/abs/1904.00221}
  {arXiv:1904.00221 [hep-ph]} \BibitemShut {NoStop}%
%%CITATION = ARXIV:1904.00221;%%
\bibitem [{\citenamefont {Shimizu}\ \emph {et~al.}(2019)\citenamefont
  {Shimizu}, \citenamefont {Yamaguchi},\ and\ \citenamefont
  {Harada}}]{Shimizu:2019ptd}%
  \BibitemOpen
  \bibfield  {author} {\bibinfo {author} {\bibfnamefont {Y.}~\bibnamefont
  {Shimizu}}, \bibinfo {author} {\bibfnamefont {Y.}~\bibnamefont {Yamaguchi}},
  \ and\ \bibinfo {author} {\bibfnamefont {M.}~\bibnamefont {Harada}},\
  }\href@noop {} {\  (\bibinfo {year} {2019})},\ \Eprint
  {http://arxiv.org/abs/1904.00587} {arXiv:1904.00587 [hep-ph]} \BibitemShut
  {NoStop}%
%%CITATION = ARXIV:1904.00587;%%
\bibitem [{\citenamefont {Guo}\ and\ \citenamefont
  {Oller}(2019)}]{Guo:2019kdc}%
  \BibitemOpen
  \bibfield  {author} {\bibinfo {author} {\bibfnamefont {Z.-H.}\ \bibnamefont
  {Guo}}\ and\ \bibinfo {author} {\bibfnamefont {J.~A.}\ \bibnamefont
  {Oller}},\ }\href {\doibase 10.1016/j.physletb.2019.04.053} {\bibfield
  {journal} {\bibinfo  {journal} {Phys. Lett.}\ }\textbf {\bibinfo {volume}
  {B793}},\ \bibinfo {pages} {144} (\bibinfo {year} {2019})},\ \Eprint
  {http://arxiv.org/abs/1904.00851} {arXiv:1904.00851 [hep-ph]} \BibitemShut
  {NoStop}%
%%CITATION = ARXIV:1904.00851;%%
\bibitem [{\citenamefont {Xiao}\ \emph
  {et~al.}(2019{\natexlab{a}})\citenamefont {Xiao}, \citenamefont {Nieves},\
  and\ \citenamefont {Oset}}]{Xiao:2019aya}%
  \BibitemOpen
  \bibfield  {author} {\bibinfo {author} {\bibfnamefont {C.~W.}\ \bibnamefont
  {Xiao}}, \bibinfo {author} {\bibfnamefont {J.}~\bibnamefont {Nieves}}, \ and\
  \bibinfo {author} {\bibfnamefont {E.}~\bibnamefont {Oset}},\ }\href {\doibase
  10.1103/PhysRevD.100.014021} {\bibfield  {journal} {\bibinfo  {journal}
  {Phys. Rev.}\ }\textbf {\bibinfo {volume} {D100}},\ \bibinfo {pages} {014021}
  (\bibinfo {year} {2019}{\natexlab{a}})},\ \Eprint
  {http://arxiv.org/abs/1904.01296} {arXiv:1904.01296 [hep-ph]} \BibitemShut
  {NoStop}%
%%CITATION = ARXIV:1904.01296;%%
\bibitem [{\citenamefont {Xiao}\ \emph
  {et~al.}(2019{\natexlab{b}})\citenamefont {Xiao}, \citenamefont {Huang},
  \citenamefont {Dong}, \citenamefont {Geng},\ and\ \citenamefont
  {Chen}}]{Xiao:2019mst}%
  \BibitemOpen
  \bibfield  {author} {\bibinfo {author} {\bibfnamefont {C.-J.}\ \bibnamefont
  {Xiao}}, \bibinfo {author} {\bibfnamefont {Y.}~\bibnamefont {Huang}},
  \bibinfo {author} {\bibfnamefont {Y.-B.}\ \bibnamefont {Dong}}, \bibinfo
  {author} {\bibfnamefont {L.-S.}\ \bibnamefont {Geng}}, \ and\ \bibinfo
  {author} {\bibfnamefont {D.-Y.}\ \bibnamefont {Chen}},\ }\href {\doibase
  10.1103/PhysRevD.100.014022} {\bibfield  {journal} {\bibinfo  {journal}
  {Phys. Rev.}\ }\textbf {\bibinfo {volume} {D100}},\ \bibinfo {pages} {014022}
  (\bibinfo {year} {2019}{\natexlab{b}})},\ \Eprint
  {http://arxiv.org/abs/1904.00872} {arXiv:1904.00872 [hep-ph]} \BibitemShut
  {NoStop}%
%%CITATION = ARXIV:1904.00872;%%
\bibitem [{\citenamefont {Sakai}\ \emph {et~al.}(2019)\citenamefont {Sakai},
  \citenamefont {Jing},\ and\ \citenamefont {Guo}}]{Sakai:2019qph}%
  \BibitemOpen
  \bibfield  {author} {\bibinfo {author} {\bibfnamefont {S.}~\bibnamefont
  {Sakai}}, \bibinfo {author} {\bibfnamefont {H.-J.}\ \bibnamefont {Jing}}, \
  and\ \bibinfo {author} {\bibfnamefont {F.-K.}\ \bibnamefont {Guo}},\
  }\href@noop {} {\  (\bibinfo {year} {2019})},\ \Eprint
  {http://arxiv.org/abs/1907.03414} {arXiv:1907.03414 [hep-ph]} \BibitemShut
  {NoStop}%
%%CITATION = ARXIV:1907.03414;%%
\bibitem [{\citenamefont {Chen}\ \emph {et~al.}(2016)\citenamefont {Chen},
  \citenamefont {Chen}, \citenamefont {Liu},\ and\ \citenamefont
  {Zhu}}]{Chen:2016qju}%
  \BibitemOpen
  \bibfield  {author} {\bibinfo {author} {\bibfnamefont {H.-X.}\ \bibnamefont
  {Chen}}, \bibinfo {author} {\bibfnamefont {W.}~\bibnamefont {Chen}}, \bibinfo
  {author} {\bibfnamefont {X.}~\bibnamefont {Liu}}, \ and\ \bibinfo {author}
  {\bibfnamefont {S.-L.}\ \bibnamefont {Zhu}},\ }\href {\doibase
  10.1016/j.physrep.2016.05.004} {\bibfield  {journal} {\bibinfo  {journal}
  {Phys. Rept.}\ }\textbf {\bibinfo {volume} {639}},\ \bibinfo {pages} {1}
  (\bibinfo {year} {2016})},\ \Eprint {http://arxiv.org/abs/1601.02092}
  {arXiv:1601.02092 [hep-ph]} \BibitemShut {NoStop}%
%%CITATION = ARXIV:1601.02092;%%
\bibitem [{\citenamefont {Guo}\ \emph {et~al.}(2018)\citenamefont {Guo},
  \citenamefont {Hanhart}, \citenamefont {Meißner}, \citenamefont {Wang},
  \citenamefont {Zhao},\ and\ \citenamefont {Zou}}]{Guo:2017jvc}%
  \BibitemOpen
  \bibfield  {author} {\bibinfo {author} {\bibfnamefont {F.-K.}\ \bibnamefont
  {Guo}}, \bibinfo {author} {\bibfnamefont {C.}~\bibnamefont {Hanhart}},
  \bibinfo {author} {\bibfnamefont {U.-G.}\ \bibnamefont {Meißner}}, \bibinfo
  {author} {\bibfnamefont {Q.}~\bibnamefont {Wang}}, \bibinfo {author}
  {\bibfnamefont {Q.}~\bibnamefont {Zhao}}, \ and\ \bibinfo {author}
  {\bibfnamefont {B.-S.}\ \bibnamefont {Zou}},\ }\href {\doibase
  10.1103/RevModPhys.90.015004} {\bibfield  {journal} {\bibinfo  {journal}
  {Rev. Mod. Phys.}\ }\textbf {\bibinfo {volume} {90}},\ \bibinfo {pages}
  {015004} (\bibinfo {year} {2018})},\ \Eprint
  {http://arxiv.org/abs/1705.00141} {arXiv:1705.00141 [hep-ph]} \BibitemShut
  {NoStop}%
%%CITATION = ARXIV:1705.00141;%%
\bibitem [{\citenamefont {Zou}\ and\ \citenamefont
  {Hussain}(2003)}]{Zou:2002yy}%
  \BibitemOpen
  \bibfield  {author} {\bibinfo {author} {\bibfnamefont {B.~S.}\ \bibnamefont
  {Zou}}\ and\ \bibinfo {author} {\bibfnamefont {F.}~\bibnamefont {Hussain}},\
  }\href {\doibase 10.1103/PhysRevC.67.015204} {\bibfield  {journal} {\bibinfo
  {journal} {Phys. Rev.}\ }\textbf {\bibinfo {volume} {C67}},\ \bibinfo {pages}
  {015204} (\bibinfo {year} {2003})},\ \Eprint
  {http://arxiv.org/abs/hep-ph/0210164} {arXiv:hep-ph/0210164 [hep-ph]}
  \BibitemShut {NoStop}%
%%CITATION = HEP-PH/0210164;%%
\bibitem [{\citenamefont {Weinberg}(1963)}]{Weinberg:1962hj}%
  \BibitemOpen
  \bibfield  {author} {\bibinfo {author} {\bibfnamefont {S.}~\bibnamefont
  {Weinberg}},\ }\href {\doibase 10.1103/PhysRev.130.776} {\bibfield  {journal}
  {\bibinfo  {journal} {Phys. Rev.}\ }\textbf {\bibinfo {volume} {130}},\
  \bibinfo {pages} {776} (\bibinfo {year} {1963})}\BibitemShut {NoStop}%
%%CITATION = PHRVA,130,776;%%
\bibitem [{\citenamefont {Weinberg}(1965)}]{Weinberg:1965zz}%
  \BibitemOpen
  \bibfield  {author} {\bibinfo {author} {\bibfnamefont {S.}~\bibnamefont
  {Weinberg}},\ }\href {\doibase 10.1103/PhysRev.137.B672} {\bibfield
  {journal} {\bibinfo  {journal} {Phys. Rev.}\ }\textbf {\bibinfo {volume}
  {137}},\ \bibinfo {pages} {B672} (\bibinfo {year} {1965})}\BibitemShut
  {NoStop}%
%%CITATION = PHRVA,137,B672;%%
\bibitem [{\citenamefont {de~Swart}(1963)}]{deSwart:1963pdg}%
  \BibitemOpen
  \bibfield  {author} {\bibinfo {author} {\bibfnamefont {J.~J.}\ \bibnamefont
  {de~Swart}},\ }\href {\doibase 10.1103/RevModPhys.35.916} {\bibfield
  {journal} {\bibinfo  {journal} {Rev. Mod. Phys.}\ }\textbf {\bibinfo {volume}
  {35}},\ \bibinfo {pages} {916} (\bibinfo {year} {1963})},\ \bibinfo {note}
  {[Erratum: Rev. Mod. Phys.37,326(1965)]}\BibitemShut {NoStop}%
%%CITATION = RMPHA,35,916;%%
\bibitem [{\citenamefont {Polinder}\ \emph {et~al.}(2006)\citenamefont
  {Polinder}, \citenamefont {Haidenbauer},\ and\ \citenamefont
  {Meissner}}]{Polinder:2006zh}%
  \BibitemOpen
  \bibfield  {author} {\bibinfo {author} {\bibfnamefont {H.}~\bibnamefont
  {Polinder}}, \bibinfo {author} {\bibfnamefont {J.}~\bibnamefont
  {Haidenbauer}}, \ and\ \bibinfo {author} {\bibfnamefont {U.-G.}\ \bibnamefont
  {Meissner}},\ }\href {\doibase 10.1016/j.nuclphysa.2006.09.006} {\bibfield
  {journal} {\bibinfo  {journal} {Nucl. Phys.}\ }\textbf {\bibinfo {volume}
  {A779}},\ \bibinfo {pages} {244} (\bibinfo {year} {2006})},\ \Eprint
  {http://arxiv.org/abs/nucl-th/0605050} {arXiv:nucl-th/0605050 [nucl-th]}
  \BibitemShut {NoStop}%
%%CITATION = NUCL-TH/0605050;%%
\bibitem [{\citenamefont {Ronchen}\ \emph {et~al.}(2013)\citenamefont
  {Ronchen}, \citenamefont {Doring}, \citenamefont {Huang}, \citenamefont
  {Haberzettl}, \citenamefont {Haidenbauer}, \citenamefont {Hanhart},
  \citenamefont {Krewald}, \citenamefont {Meissner},\ and\ \citenamefont
  {Nakayama}}]{Ronchen:2012eg}%
  \BibitemOpen
  \bibfield  {author} {\bibinfo {author} {\bibfnamefont {D.}~\bibnamefont
  {Ronchen}}, \bibinfo {author} {\bibfnamefont {M.}~\bibnamefont {Doring}},
  \bibinfo {author} {\bibfnamefont {F.}~\bibnamefont {Huang}}, \bibinfo
  {author} {\bibfnamefont {H.}~\bibnamefont {Haberzettl}}, \bibinfo {author}
  {\bibfnamefont {J.}~\bibnamefont {Haidenbauer}}, \bibinfo {author}
  {\bibfnamefont {C.}~\bibnamefont {Hanhart}}, \bibinfo {author} {\bibfnamefont
  {S.}~\bibnamefont {Krewald}}, \bibinfo {author} {\bibfnamefont {U.~G.}\
  \bibnamefont {Meissner}}, \ and\ \bibinfo {author} {\bibfnamefont
  {K.}~\bibnamefont {Nakayama}},\ }\href {\doibase 10.1140/epja/i2013-13044-5}
  {\bibfield  {journal} {\bibinfo  {journal} {Eur. Phys. J.}\ }\textbf
  {\bibinfo {volume} {A49}},\ \bibinfo {pages} {44} (\bibinfo {year} {2013})},\
  \Eprint {http://arxiv.org/abs/1211.6998} {arXiv:1211.6998 [nucl-th]}
  \BibitemShut {NoStop}%
%%CITATION = ARXIV:1211.6998;%%
\bibitem [{\citenamefont {Haidenbauer}\ and\ \citenamefont
  {Krein}(2017)}]{Haidenbauer:2016pva}%
  \BibitemOpen
  \bibfield  {author} {\bibinfo {author} {\bibfnamefont {J.}~\bibnamefont
  {Haidenbauer}}\ and\ \bibinfo {author} {\bibfnamefont {G.}~\bibnamefont
  {Krein}},\ }\href {\doibase 10.1103/PhysRevD.95.014017} {\bibfield  {journal}
  {\bibinfo  {journal} {Phys. Rev.}\ }\textbf {\bibinfo {volume} {D95}},\
  \bibinfo {pages} {014017} (\bibinfo {year} {2017})},\ \Eprint
  {http://arxiv.org/abs/1611.02985} {arXiv:1611.02985 [nucl-th]} \BibitemShut
  {NoStop}%
%%CITATION = ARXIV:1611.02985;%%
\bibitem [{\citenamefont {Haidenbauer}\ \emph {et~al.}(2017)\citenamefont
  {Haidenbauer}, \citenamefont {Petschauer}, \citenamefont {Kaiser},
  \citenamefont {Meißner},\ and\ \citenamefont {Weise}}]{Haidenbauer:2017sws}%
  \BibitemOpen
  \bibfield  {author} {\bibinfo {author} {\bibfnamefont {J.}~\bibnamefont
  {Haidenbauer}}, \bibinfo {author} {\bibfnamefont {S.}~\bibnamefont
  {Petschauer}}, \bibinfo {author} {\bibfnamefont {N.}~\bibnamefont {Kaiser}},
  \bibinfo {author} {\bibfnamefont {U.-G.}\ \bibnamefont {Meißner}}, \ and\
  \bibinfo {author} {\bibfnamefont {W.}~\bibnamefont {Weise}},\ }\href
  {\doibase 10.1140/epjc/s10052-017-5309-4} {\bibfield  {journal} {\bibinfo
  {journal} {Eur. Phys. J.}\ }\textbf {\bibinfo {volume} {C77}},\ \bibinfo
  {pages} {760} (\bibinfo {year} {2017})},\ \Eprint
  {http://arxiv.org/abs/1708.08071} {arXiv:1708.08071 [nucl-th]} \BibitemShut
  {NoStop}%
%%CITATION = ARXIV:1708.08071;%%
\bibitem [{\citenamefont {Colangelo}\ \emph {et~al.}(2004)\citenamefont
  {Colangelo}, \citenamefont {De~Fazio},\ and\ \citenamefont
  {Pham}}]{Colangelo:2003sa}%
  \BibitemOpen
  \bibfield  {author} {\bibinfo {author} {\bibfnamefont {P.}~\bibnamefont
  {Colangelo}}, \bibinfo {author} {\bibfnamefont {F.}~\bibnamefont {De~Fazio}},
  \ and\ \bibinfo {author} {\bibfnamefont {T.~N.}\ \bibnamefont {Pham}},\
  }\href {\doibase 10.1103/PhysRevD.69.054023} {\bibfield  {journal} {\bibinfo
  {journal} {Phys. Rev.}\ }\textbf {\bibinfo {volume} {D69}},\ \bibinfo {pages}
  {054023} (\bibinfo {year} {2004})},\ \Eprint
  {http://arxiv.org/abs/hep-ph/0310084} {arXiv:hep-ph/0310084 [hep-ph]}
  \BibitemShut {NoStop}%
%%CITATION = HEP-PH/0310084;%%
\bibitem [{\citenamefont {Guo}\ \emph {et~al.}(2011)\citenamefont {Guo},
  \citenamefont {Hanhart}, \citenamefont {Li}, \citenamefont {Meissner},\ and\
  \citenamefont {Zhao}}]{Guo:2010ak}%
  \BibitemOpen
  \bibfield  {author} {\bibinfo {author} {\bibfnamefont {F.-K.}\ \bibnamefont
  {Guo}}, \bibinfo {author} {\bibfnamefont {C.}~\bibnamefont {Hanhart}},
  \bibinfo {author} {\bibfnamefont {G.}~\bibnamefont {Li}}, \bibinfo {author}
  {\bibfnamefont {U.-G.}\ \bibnamefont {Meissner}}, \ and\ \bibinfo {author}
  {\bibfnamefont {Q.}~\bibnamefont {Zhao}},\ }\href {\doibase
  10.1103/PhysRevD.83.034013} {\bibfield  {journal} {\bibinfo  {journal} {Phys.
  Rev.}\ }\textbf {\bibinfo {volume} {D83}},\ \bibinfo {pages} {034013}
  (\bibinfo {year} {2011})},\ \Eprint {http://arxiv.org/abs/1008.3632}
  {arXiv:1008.3632 [hep-ph]} \BibitemShut {NoStop}%
%%CITATION = ARXIV:1008.3632;%%
\bibitem [{\citenamefont {Lin}\ and\ \citenamefont {Ko}(2000)}]{Lin:1999ad}%
  \BibitemOpen
  \bibfield  {author} {\bibinfo {author} {\bibfnamefont {Z.-w.}\ \bibnamefont
  {Lin}}\ and\ \bibinfo {author} {\bibfnamefont {C.~M.}\ \bibnamefont {Ko}},\
  }\href {\doibase 10.1103/PhysRevC.62.034903} {\bibfield  {journal} {\bibinfo
  {journal} {Phys. Rev.}\ }\textbf {\bibinfo {volume} {C62}},\ \bibinfo {pages}
  {034903} (\bibinfo {year} {2000})},\ \Eprint
  {http://arxiv.org/abs/nucl-th/9912046} {arXiv:nucl-th/9912046 [nucl-th]}
  \BibitemShut {NoStop}%
%%CITATION = NUCL-TH/9912046;%%
\bibitem [{\citenamefont {Oh}\ \emph {et~al.}(2001)\citenamefont {Oh},
  \citenamefont {Song},\ and\ \citenamefont {Lee}}]{Oh:2000qr}%
  \BibitemOpen
  \bibfield  {author} {\bibinfo {author} {\bibfnamefont {Y.-s.}\ \bibnamefont
  {Oh}}, \bibinfo {author} {\bibfnamefont {T.}~\bibnamefont {Song}}, \ and\
  \bibinfo {author} {\bibfnamefont {S.~H.}\ \bibnamefont {Lee}},\ }\href
  {\doibase 10.1103/PhysRevC.63.034901} {\bibfield  {journal} {\bibinfo
  {journal} {Phys. Rev.}\ }\textbf {\bibinfo {volume} {C63}},\ \bibinfo {pages}
  {034901} (\bibinfo {year} {2001})},\ \Eprint
  {http://arxiv.org/abs/nucl-th/0010064} {arXiv:nucl-th/0010064 [nucl-th]}
  \BibitemShut {NoStop}%
%%CITATION = NUCL-TH/0010064;%%
\bibitem [{\citenamefont {Yan}\ \emph {et~al.}(1992)\citenamefont {Yan},
  \citenamefont {Cheng}, \citenamefont {Cheung}, \citenamefont {Lin},
  \citenamefont {Lin},\ and\ \citenamefont {Yu}}]{Yan:1992gz}%
  \BibitemOpen
  \bibfield  {author} {\bibinfo {author} {\bibfnamefont {T.-M.}\ \bibnamefont
  {Yan}}, \bibinfo {author} {\bibfnamefont {H.-Y.}\ \bibnamefont {Cheng}},
  \bibinfo {author} {\bibfnamefont {C.-Y.}\ \bibnamefont {Cheung}}, \bibinfo
  {author} {\bibfnamefont {G.-L.}\ \bibnamefont {Lin}}, \bibinfo {author}
  {\bibfnamefont {Y.~C.}\ \bibnamefont {Lin}}, \ and\ \bibinfo {author}
  {\bibfnamefont {H.-L.}\ \bibnamefont {Yu}},\ }\href {\doibase
  10.1103/PhysRevD.46.1148, 10.1103/PhysRevD.55.5851} {\bibfield  {journal}
  {\bibinfo  {journal} {Phys. Rev.}\ }\textbf {\bibinfo {volume} {D46}},\
  \bibinfo {pages} {1148} (\bibinfo {year} {1992})},\ \bibinfo {note}
  {[Erratum: Phys. Rev.D55,5851(1997)]}\BibitemShut {NoStop}%
%%CITATION = PHRVA,D46,1148;%%
\bibitem [{\citenamefont {Cheng}\ \emph {et~al.}(2005)\citenamefont {Cheng},
  \citenamefont {Chua},\ and\ \citenamefont {Soni}}]{Cheng:2004ru}%
  \BibitemOpen
  \bibfield  {author} {\bibinfo {author} {\bibfnamefont {H.-Y.}\ \bibnamefont
  {Cheng}}, \bibinfo {author} {\bibfnamefont {C.-K.}\ \bibnamefont {Chua}}, \
  and\ \bibinfo {author} {\bibfnamefont {A.}~\bibnamefont {Soni}},\ }\href
  {\doibase 10.1103/PhysRevD.71.014030} {\bibfield  {journal} {\bibinfo
  {journal} {Phys. Rev.}\ }\textbf {\bibinfo {volume} {D71}},\ \bibinfo {pages}
  {014030} (\bibinfo {year} {2005})},\ \Eprint
  {http://arxiv.org/abs/hep-ph/0409317} {arXiv:hep-ph/0409317 [hep-ph]}
  \BibitemShut {NoStop}%
%%CITATION = HEP-PH/0409317;%%
\bibitem [{\citenamefont {Albaladejo}\ \emph {et~al.}(2015)\citenamefont
  {Albaladejo}, \citenamefont {Guo}, \citenamefont {Hidalgo-Duque},
  \citenamefont {Nieves},\ and\ \citenamefont
  {Valderrama}}]{Albaladejo:2015dsa}%
  \BibitemOpen
  \bibfield  {author} {\bibinfo {author} {\bibfnamefont {M.}~\bibnamefont
  {Albaladejo}}, \bibinfo {author} {\bibfnamefont {F.~K.}\ \bibnamefont {Guo}},
  \bibinfo {author} {\bibfnamefont {C.}~\bibnamefont {Hidalgo-Duque}}, \bibinfo
  {author} {\bibfnamefont {J.}~\bibnamefont {Nieves}}, \ and\ \bibinfo {author}
  {\bibfnamefont {M.~P.}\ \bibnamefont {Valderrama}},\ }\href {\doibase
  10.1140/epjc/s10052-015-3753-6} {\bibfield  {journal} {\bibinfo  {journal}
  {Eur. Phys. J.}\ }\textbf {\bibinfo {volume} {C75}},\ \bibinfo {pages} {547}
  (\bibinfo {year} {2015})},\ \Eprint {http://arxiv.org/abs/1504.00861}
  {arXiv:1504.00861 [hep-ph]} \BibitemShut {NoStop}%
%%CITATION = ARXIV:1504.00861;%%
\bibitem [{\citenamefont {Shen}\ \emph {et~al.}(2016)\citenamefont {Shen},
  \citenamefont {Guo}, \citenamefont {Xie},\ and\ \citenamefont
  {Zou}}]{Shen:2016tzq}%
  \BibitemOpen
  \bibfield  {author} {\bibinfo {author} {\bibfnamefont {C.-W.}\ \bibnamefont
  {Shen}}, \bibinfo {author} {\bibfnamefont {F.-K.}\ \bibnamefont {Guo}},
  \bibinfo {author} {\bibfnamefont {J.-J.}\ \bibnamefont {Xie}}, \ and\
  \bibinfo {author} {\bibfnamefont {B.-S.}\ \bibnamefont {Zou}},\ }\href
  {\doibase 10.1016/j.nuclphysa.2016.04.034} {\bibfield  {journal} {\bibinfo
  {journal} {Nucl. Phys.}\ }\textbf {\bibinfo {volume} {A954}},\ \bibinfo
  {pages} {393} (\bibinfo {year} {2016})},\ \Eprint
  {http://arxiv.org/abs/1603.04672} {arXiv:1603.04672 [hep-ph]} \BibitemShut
  {NoStop}%
%%CITATION = ARXIV:1603.04672;%%
\bibitem [{\citenamefont {Faessler}\ \emph {et~al.}(2007)\citenamefont
  {Faessler}, \citenamefont {Gutsche}, \citenamefont {Lyubovitskij},\ and\
  \citenamefont {Ma}}]{Faessler:2007gv}%
  \BibitemOpen
  \bibfield  {author} {\bibinfo {author} {\bibfnamefont {A.}~\bibnamefont
  {Faessler}}, \bibinfo {author} {\bibfnamefont {T.}~\bibnamefont {Gutsche}},
  \bibinfo {author} {\bibfnamefont {V.~E.}\ \bibnamefont {Lyubovitskij}}, \
  and\ \bibinfo {author} {\bibfnamefont {Y.-L.}\ \bibnamefont {Ma}},\ }\href
  {\doibase 10.1103/PhysRevD.76.014005} {\bibfield  {journal} {\bibinfo
  {journal} {Phys. Rev.}\ }\textbf {\bibinfo {volume} {D76}},\ \bibinfo {pages}
  {014005} (\bibinfo {year} {2007})},\ \Eprint {http://arxiv.org/abs/0705.0254}
  {arXiv:0705.0254 [hep-ph]} \BibitemShut {NoStop}%
%%CITATION = ARXIV:0705.0254;%%
\bibitem [{\citenamefont {Dong}\ \emph {et~al.}(2009)\citenamefont {Dong},
  \citenamefont {Faessler}, \citenamefont {Gutsche}, \citenamefont
  {Kovalenko},\ and\ \citenamefont {Lyubovitskij}}]{Dong:2009yp}%
  \BibitemOpen
  \bibfield  {author} {\bibinfo {author} {\bibfnamefont {Y.}~\bibnamefont
  {Dong}}, \bibinfo {author} {\bibfnamefont {A.}~\bibnamefont {Faessler}},
  \bibinfo {author} {\bibfnamefont {T.}~\bibnamefont {Gutsche}}, \bibinfo
  {author} {\bibfnamefont {S.}~\bibnamefont {Kovalenko}}, \ and\ \bibinfo
  {author} {\bibfnamefont {V.~E.}\ \bibnamefont {Lyubovitskij}},\ }\href
  {\doibase 10.1103/PhysRevD.79.094013} {\bibfield  {journal} {\bibinfo
  {journal} {Phys. Rev.}\ }\textbf {\bibinfo {volume} {D79}},\ \bibinfo {pages}
  {094013} (\bibinfo {year} {2009})},\ \Eprint {http://arxiv.org/abs/0903.5416}
  {arXiv:0903.5416 [hep-ph]} \BibitemShut {NoStop}%
%%CITATION = ARXIV:0903.5416;%%
\bibitem [{\citenamefont {Dong}\ \emph {et~al.}(2010)\citenamefont {Dong},
  \citenamefont {Faessler}, \citenamefont {Gutsche},\ and\ \citenamefont
  {Lyubovitskij}}]{Dong:2009tg}%
  \BibitemOpen
  \bibfield  {author} {\bibinfo {author} {\bibfnamefont {Y.}~\bibnamefont
  {Dong}}, \bibinfo {author} {\bibfnamefont {A.}~\bibnamefont {Faessler}},
  \bibinfo {author} {\bibfnamefont {T.}~\bibnamefont {Gutsche}}, \ and\
  \bibinfo {author} {\bibfnamefont {V.~E.}\ \bibnamefont {Lyubovitskij}},\
  }\href {\doibase 10.1103/PhysRevD.81.014006} {\bibfield  {journal} {\bibinfo
  {journal} {Phys. Rev.}\ }\textbf {\bibinfo {volume} {D81}},\ \bibinfo {pages}
  {014006} (\bibinfo {year} {2010})},\ \Eprint {http://arxiv.org/abs/0910.1204}
  {arXiv:0910.1204 [hep-ph]} \BibitemShut {NoStop}%
%%CITATION = ARXIV:0910.1204;%%
\bibitem [{\citenamefont {Lü}\ and\ \citenamefont {Dong}(2016)}]{Lu:2016nnt}%
  \BibitemOpen
  \bibfield  {author} {\bibinfo {author} {\bibfnamefont {Q.-F.}\ \bibnamefont
  {Lü}}\ and\ \bibinfo {author} {\bibfnamefont {Y.-B.}\ \bibnamefont {Dong}},\
  }\href {\doibase 10.1103/PhysRevD.93.074020} {\bibfield  {journal} {\bibinfo
  {journal} {Phys. Rev.}\ }\textbf {\bibinfo {volume} {D93}},\ \bibinfo {pages}
  {074020} (\bibinfo {year} {2016})},\ \Eprint
  {http://arxiv.org/abs/1603.00559} {arXiv:1603.00559 [hep-ph]} \BibitemShut
  {NoStop}%
%%CITATION = ARXIV:1603.00559;%%
\bibitem [{\citenamefont {Nieves}\ and\ \citenamefont
  {Valderrama}(2012)}]{Nieves:2012tt}%
  \BibitemOpen
  \bibfield  {author} {\bibinfo {author} {\bibfnamefont {J.}~\bibnamefont
  {Nieves}}\ and\ \bibinfo {author} {\bibfnamefont {M.~P.}\ \bibnamefont
  {Valderrama}},\ }\href {\doibase 10.1103/PhysRevD.86.056004} {\bibfield
  {journal} {\bibinfo  {journal} {Phys. Rev.}\ }\textbf {\bibinfo {volume}
  {D86}},\ \bibinfo {pages} {056004} (\bibinfo {year} {2012})},\ \Eprint
  {http://arxiv.org/abs/1204.2790} {arXiv:1204.2790 [hep-ph]} \BibitemShut
  {NoStop}%
%%CITATION = ARXIV:1204.2790;%%
\bibitem [{\citenamefont {Hidalgo-Duque}\ \emph {et~al.}(2013)\citenamefont
  {Hidalgo-Duque}, \citenamefont {Nieves},\ and\ \citenamefont
  {Valderrama}}]{HidalgoDuque:2012pq}%
  \BibitemOpen
  \bibfield  {author} {\bibinfo {author} {\bibfnamefont {C.}~\bibnamefont
  {Hidalgo-Duque}}, \bibinfo {author} {\bibfnamefont {J.}~\bibnamefont
  {Nieves}}, \ and\ \bibinfo {author} {\bibfnamefont {M.~P.}\ \bibnamefont
  {Valderrama}},\ }\href {\doibase 10.1103/PhysRevD.87.076006} {\bibfield
  {journal} {\bibinfo  {journal} {Phys. Rev.}\ }\textbf {\bibinfo {volume}
  {D87}},\ \bibinfo {pages} {076006} (\bibinfo {year} {2013})},\ \Eprint
  {http://arxiv.org/abs/1210.5431} {arXiv:1210.5431 [hep-ph]} \BibitemShut
  {NoStop}%
%%CITATION = ARXIV:1210.5431;%%
\bibitem [{\citenamefont {Ali}\ \emph {et~al.}(2019)\citenamefont {Ali} \emph
  {et~al.}}]{Ali:2019lzf}%
  \BibitemOpen
  \bibfield  {author} {\bibinfo {author} {\bibfnamefont {A.}~\bibnamefont
  {Ali}} \emph {et~al.} (\bibinfo {collaboration} {GlueX}),\ }\href@noop {} {\
  (\bibinfo {year} {2019})},\ \Eprint {http://arxiv.org/abs/1905.10811}
  {arXiv:1905.10811 [nucl-ex]} \BibitemShut {NoStop}%
%%CITATION = ARXIV:1905.10811;%%
\bibitem [{\citenamefont {Voloshin}(2019)}]{Voloshin:2019aut}%
  \BibitemOpen
  \bibfield  {author} {\bibinfo {author} {\bibfnamefont {M.~B.}\ \bibnamefont
  {Voloshin}},\ }\href@noop {} {\  (\bibinfo {year} {2019})},\ \Eprint
  {http://arxiv.org/abs/1907.01476} {arXiv:1907.01476 [hep-ph]} \BibitemShut
  {NoStop}%
%%CITATION = ARXIV:1907.01476;%%
\bibitem [{\citenamefont {Yamaguchi}\ \emph {et~al.}(2019)\citenamefont
  {Yamaguchi}, \citenamefont {García-Tecocoatzi}, \citenamefont {Giachino},
  \citenamefont {Hosaka}, \citenamefont {Santopinto}, \citenamefont
  {Takeuchi},\ and\ \citenamefont {Takizawa}}]{Yamaguchi:2019seo}%
  \BibitemOpen
  \bibfield  {author} {\bibinfo {author} {\bibfnamefont {Y.}~\bibnamefont
  {Yamaguchi}}, \bibinfo {author} {\bibfnamefont {H.}~\bibnamefont
  {García-Tecocoatzi}}, \bibinfo {author} {\bibfnamefont {A.}~\bibnamefont
  {Giachino}}, \bibinfo {author} {\bibfnamefont {A.}~\bibnamefont {Hosaka}},
  \bibinfo {author} {\bibfnamefont {E.}~\bibnamefont {Santopinto}}, \bibinfo
  {author} {\bibfnamefont {S.}~\bibnamefont {Takeuchi}}, \ and\ \bibinfo
  {author} {\bibfnamefont {M.}~\bibnamefont {Takizawa}},\ }\href@noop {} {\
  (\bibinfo {year} {2019})},\ \Eprint {http://arxiv.org/abs/1907.04684}
  {arXiv:1907.04684 [hep-ph]} \BibitemShut {NoStop}%
%%CITATION = ARXIV:1907.04684;%%
\bibitem [{\citenamefont {Pavon~Valderrama}(2019)}]{Valderrama:2019chc}%
  \BibitemOpen
  \bibfield  {author} {\bibinfo {author} {\bibfnamefont {M.}~\bibnamefont
  {Pavon~Valderrama}},\ }\href@noop {} {\  (\bibinfo {year} {2019})},\ \Eprint
  {http://arxiv.org/abs/1907.05294} {arXiv:1907.05294 [hep-ph]} \BibitemShut
  {NoStop}%
%%CITATION = ARXIV:1907.05294;%%
\bibitem [{\citenamefont {Liu}\ \emph {et~al.}(2019{\natexlab{c}})\citenamefont
  {Liu}, \citenamefont {Wu}, \citenamefont {Sánchez~Sánchez}, \citenamefont
  {Valderrama}, \citenamefont {Geng},\ and\ \citenamefont {Xie}}]{Liu:2019zvb}%
  \BibitemOpen
  \bibfield  {author} {\bibinfo {author} {\bibfnamefont {M.-Z.}\ \bibnamefont
  {Liu}}, \bibinfo {author} {\bibfnamefont {T.-W.}\ \bibnamefont {Wu}},
  \bibinfo {author} {\bibfnamefont {M.}~\bibnamefont {Sánchez~Sánchez}},
  \bibinfo {author} {\bibfnamefont {M.~P.}\ \bibnamefont {Valderrama}},
  \bibinfo {author} {\bibfnamefont {L.-S.}\ \bibnamefont {Geng}}, \ and\
  \bibinfo {author} {\bibfnamefont {J.-J.}\ \bibnamefont {Xie}},\ }\href@noop
  {} {\  (\bibinfo {year} {2019}{\natexlab{c}})},\ \Eprint
  {http://arxiv.org/abs/1907.06093} {arXiv:1907.06093 [hep-ph]} \BibitemShut
  {NoStop}%
%%CITATION = ARXIV:1907.06093;%%
\bibitem [{\citenamefont {Pan}\ \emph {et~al.}(2019)\citenamefont {Pan},
  \citenamefont {Liu}, \citenamefont {Peng}, \citenamefont {Sánchez~Sánchez},
  \citenamefont {Geng},\ and\ \citenamefont {Valderrama}}]{Pan:2019skd}%
  \BibitemOpen
  \bibfield  {author} {\bibinfo {author} {\bibfnamefont {Y.-W.}\ \bibnamefont
  {Pan}}, \bibinfo {author} {\bibfnamefont {M.-Z.}\ \bibnamefont {Liu}},
  \bibinfo {author} {\bibfnamefont {F.-Z.}\ \bibnamefont {Peng}}, \bibinfo
  {author} {\bibfnamefont {M.}~\bibnamefont {Sánchez~Sánchez}}, \bibinfo
  {author} {\bibfnamefont {L.-S.}\ \bibnamefont {Geng}}, \ and\ \bibinfo
  {author} {\bibfnamefont {M.~P.}\ \bibnamefont {Valderrama}},\ }\href@noop {}
  {\  (\bibinfo {year} {2019})},\ \Eprint {http://arxiv.org/abs/1907.11220}
  {arXiv:1907.11220 [hep-ph]} \BibitemShut {NoStop}%
%%CITATION = ARXIV:1907.11220;%%
\end{thebibliography}%

\end{document}